# Heterogeneous networks in drug-target interaction prediction

Mohammad Molaee[1], Nasrollah Moghadam Charkari[1], Foad Ghaderi[1]
1 Department of Electrical and Computer Engineering, Tarbiat Modares University, Tehran, Iran
Corresponding author: Nasrollah Moghaddam Charkari (charkari@modares.ac.ir)
Contributing author: Mohammad Molaee (mohammad_molaie@modares.ac.ir), Foad Ghaderi (fghaderi@modares.ac.ir)

**Abstract**
Drug discovery requires a tremendous amount of time and cost. Computational drug-target interaction prediction, an important part of this process, can reduce these requirements by narrowing the search space for wet lab experiments. In this survey, we provide comprehensive details of graph machine learning-based methods in predicting drug-target interaction, as they have shown promising results in this field. These details include the overall framework, main contribution, datasets, and their source codes. The selected papers were mainly published from 2020 to 2024. Prior to discussing papers, we briefly introduce the datasets commonly used with these methods and measurements to assess their performance. Finally, future challenges and some crucial areas that need to be explored are discussed.

## Introduction

Drug discovery is the process of finding an effective substance to manage and treat a disease and includes research on animals, cell models, and human beings [1]. The development of a drug typically takes nine to twelve years from when it is discovered to when it becomes a commercially available drug, with only one compound out of 5,000 entering pre-clinical testing being approved [2,3]. The entire process requires an average of $900 million to $2 billion [4].

The above-mentioned problems have attracted scientists to drug repurposing, using approved and safe medications for the treatment of different illnesses [5]. For instance, minoxidil was approved for arterial hypertension but was later found to be a suitable solution for hair loss treatment [6]. This phenomenon is based on the fact that drugs may interact with off-target proteins [7,8]. Drug repurposing can decrease the cost and time wasted compared with traditional methods [9]. One of the main problems of repurposing methods is the scarcity of available data on drug-protein interactions [10].

Data in this field has increased owing to the Human Genome Project. Since this project, several large-scale databases, new technologies, and analytical tools have emerged [11], significantly enhancing the computational methods developed for drug discovery and repurposing.

In drug discovery and repurposing, predicting drug-target interaction (DTI) is a vital step [12,13]. Two main methods are employed to achieve this goal: binary classification and regression analysis. The former, which has been widely used in comparison to the latter [14], considers the task as the prediction of the existence of a binding between drug-target pairs (DTPs). Since these datasets are incomplete, meaning numerous DTIs have yet to be discovered, and these methods cannot distinguish between true negative DTIs and missing information, utilizing this method affects the performance of the models; therefore, it is more meaningful and practical to use the other approach. The regression task aims to predict the drug-target binding affinity, in other words, the strength of binding between DTPs [15].

In recent years, owing to the prospering graph machine learning, there has been significant interest in using these methods for DTI prediction. In this regard, this article focuses on different approaches in machine learning methods used for extracting knowledge from graphs, especially heterogeneous information networks, which are also known as heterogeneous networks.

In these studies, the chemical structures of drugs are commonly represented as Simplified Molecular Input Line Entry Specification (SMILES), and proteins are represented by their amino acid sequences. These representations are string-based, making them computationally more efficient than 2D methods. There are several methods, such as word2vec [16], that can be adopted for representation learning. However, some studies, which are discussed in depth, have converted SMILESs and sequences to corresponding graph representations. Table *1* shows the SMILES and molecular graphs of the two drugs collected from DrugBank [17] (https://go.drugbank.com).

Table 1 Two drugs with corresponding molecular graph representation

| Drug Name | SMILES | Molecular Graph |
|---|---|---|
| **Acetaminophen** | CC(=O)NC1=CC=C(O)C=C1 | 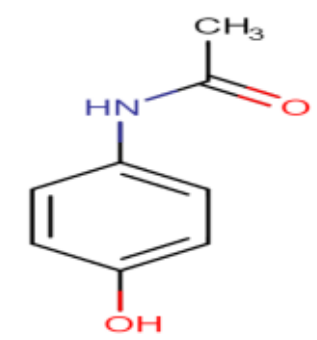 |
| **Aspirin** | CC(=O)OC1=CC=CC=C1C(O)=O | 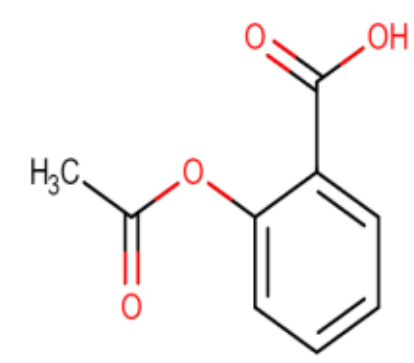 |

# Preliminaries

## Heterogeneous network

A heterogeneous information network (HIN) or heterogeneous network (HN) is defined as a directed graph G = (V, E), accompanied by a node-type mapping function τ: V → A and a link-type mapping function φ: E → R. In these networks, each node v ∈ V is associated with a specific node type τ (v) ∈ A, and each link e ∈ E is associated with a particular relation φ(e) ∈ R. If two links are assigned to the same relation type, they share identical starting- and ending-node types. When |A| + |R| > 2, the network is called a heterogeneous network [18]. Figure 1 illustrates the heterogeneous network of Luo's dataset.

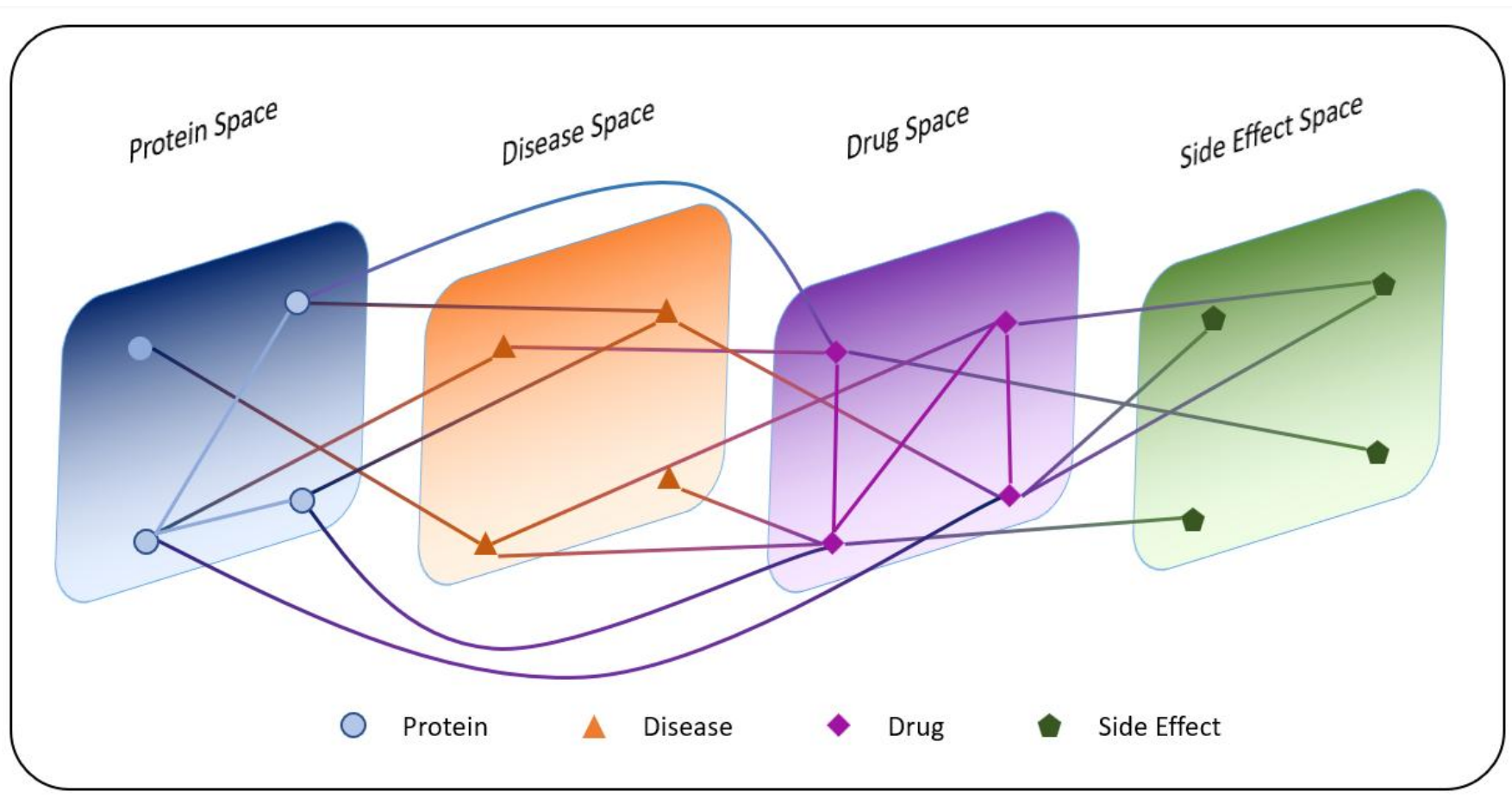

Figure 1 Overall view of a heterogeneous network, comprising four node types—protein, disease, drug, and side effect—and six link types: protein-protein, protein-disease, protein-drug, drug-disease, drug-drug, and drug-side effect. A distinct color represents each node and edge type to demonstrate the complex interconnections of the network.

## Metapath

A metapath M is defined as a path in the form of $N_1 \xrightarrow{E_1} N_2 \xrightarrow{E_2} \ldots \xrightarrow{E_{n-1}} N_n$ , where $R = E_1 \circ E_2 \circ \cdots \circ E_{n-1}$ indicates a relation between $N_1$ and $N_n$ composed of any edge type, and ∘ denotes the composition operator on edges or relations. A metapath reveals structural correlations and semantics in a heterogeneous graph [19].

## Metapath instance

Considering a metapath M, multiple metapath instances M might exist for each node, following the schema of M.

## Metapath-based neighbors

Considering metapath M, the metapath-based neighbors $S_N^M$ of node N are a multiset of nodes, including node N itself and all the other nodes connected to this node via metapath instances of M. Multiset means that if a node such as $N_i$ connects to node $N_j$ through k different metapath instances, it will be considered k times.

## Metagraph

Regarding a metapath M, all metapath-based neighbors of node N create the metagraph $G_N^M$, which is a subgraph of the main graph G. All intermediate nodes on each metapath instance will be considered as an edge in this metagraph. Figure 2 illustrates metapath, metapath instance, metapath-based neighbors, and metagraph by examples.

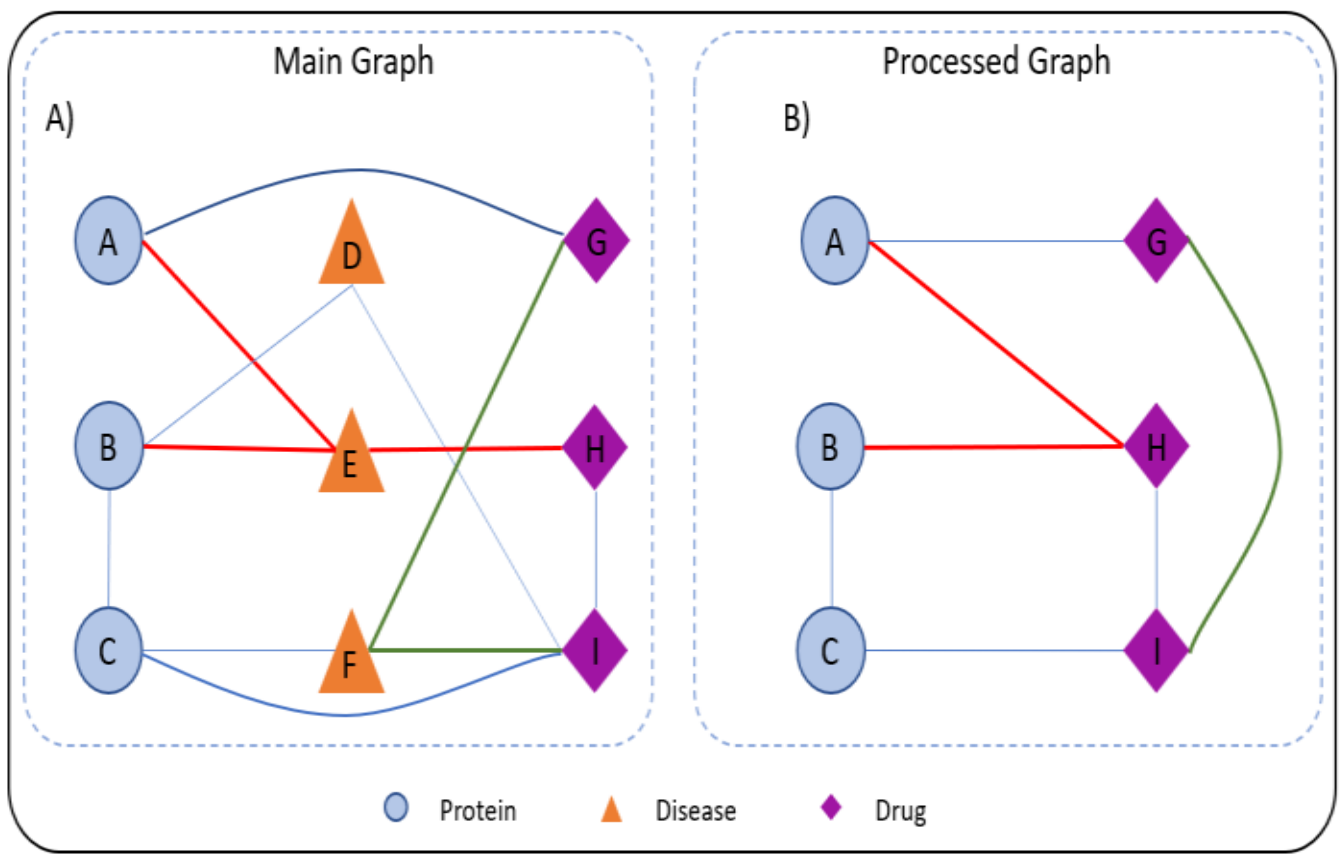

Figure 2 A) The main graph depicts two metapaths. Red metapath instances demonstrate protein-disease-drug metapath, proteins to drugs, while the green one shows drug-disease-drug metapath, connecting drugs to drugs. B) The processed graph is the metagraph acquired by removing the disease nodes and edges connected to them, which connects the two ends of the three metapath instances (metapath-based neighbors).

## Node2Vec

Node2Vec [20] learns a task-independent embedding for the nodes. The ultimate goal of this method is to find a matrix $Z \in R^{d \times |V|}$, where column i shows the representation of node $N_i$. If nodes u and v are similar in the graph space, $z_u$ and $z_v$ must be close in the representation space. To define the similarity, a biased second-order random walk, illustrated in **Error! Reference source not found.**, is employed. If node u is visited in a random walk starting from node v, they are considered similar.

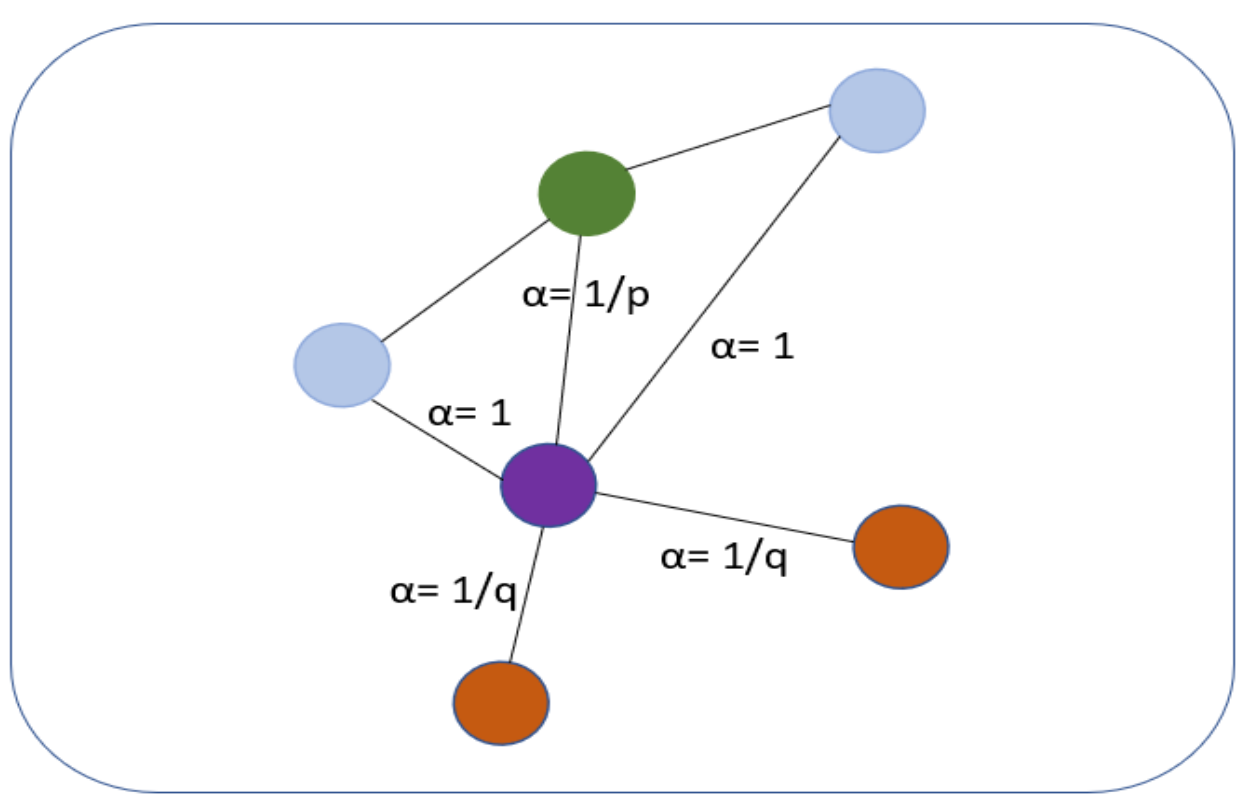

Figure 3 Edge labels (α values) denote the unnormalized transition probabilities to neighboring nodes based on the return parameter p and the in-out parameter q. Given that the walk has just transitioned from the green node to the purple node, the light blue node.

## Graph Neural Networks (GNN)

Since graphs have no structure, regular deep neural networks cannot be applied. Not only do these methods use the initial features of graph elements, such as nodes or links, but they are also task-specific; that is, they can take advantage of labels [21]. GNN models comprise two functions: messaging and aggregation. In other words, the process of calculating a new representation of a node involves aggregating the messages received from the neighbors of that node. Messaging indicates how the information of each neighbor node is sent to the node for which a new representation is being learned, while aggregation means how the received information should be combined. The aggregation function ought to be permutation-invariant, as there is no order or structure in the graphs. The key distinction among these models is how they aggregate information about the neighbors across the layers. The three most popular GNN models are briefly discussed. A graph and overview of the corresponding GNN with two layers are shown in **Error! Reference source not found.**.

**Graph Convolutional Network (GCN)**

GCN [22] uses the neighborhood aggregation and messaging function shown in Equation (1) in each layer, where A is the adjacency matrix, $\widetilde{A} = A + I_N$, $\widetilde{D}_{ii} = \sum_j \widetilde{A}_{ij}$, and $W^{(l)}$ and $H^{(l)}$ are the trainable weights and learned representation in layer l, respectively, considering $H^{(0)} = X$.

$$H^{(l+1)} = \sigma\left(\widetilde{D}^{-\frac{1}{2}}\widetilde{A}\widetilde{D}^{-\frac{1}{2}} H^{(l)}W^{(l)}\right) \quad (1)$$

Although this might seem difficult to understand, it only demonstrates that $H^{(l+1)}$ is the average (aggregation function) of the features of the node itself and its neighbors multiplied by the network weight (messaging function).

**GraphSAGE**

The embedding generation of GraphSAGE [23] follows Equations (2) and (3):

$$h_{N(v)}^{k} = \mathrm{AGGREGATE}_K\left(\left\{h_u^{k-1}, \forall\, u \in N(v)\right\}\right) \quad (2)$$

$$h_v^k = \sigma\left(\mathbf{W}^k.\mathrm{CONCAT}\left(h_v^{k-1}, h_{N(v)}^k\right)\right) \quad (3)$$

The main difference between GraphSAGE and GCN is that in this function, other functions, including pooling and LSTM aggregator, are utilized instead of the mean aggregator. Moreover, the current representation of the node itself is concatenated with the neighbor function, which works as a simple form of a skip connection.

**Graph Attention Network (GAT)**

The only difference between GAT [24] and GCN is that, as Equation (4) indicates, the weighted summation of neighbor node features is employed as an alternative to averaging, where the weight is learned for each node (attention weight).

$$h_v^k = \sigma\left(\sum\nolimits_{u\in N(v)} \alpha_{vu}\mathbf{W}^{(l)}h_u^{(l-1)}\right) \quad (4)$$

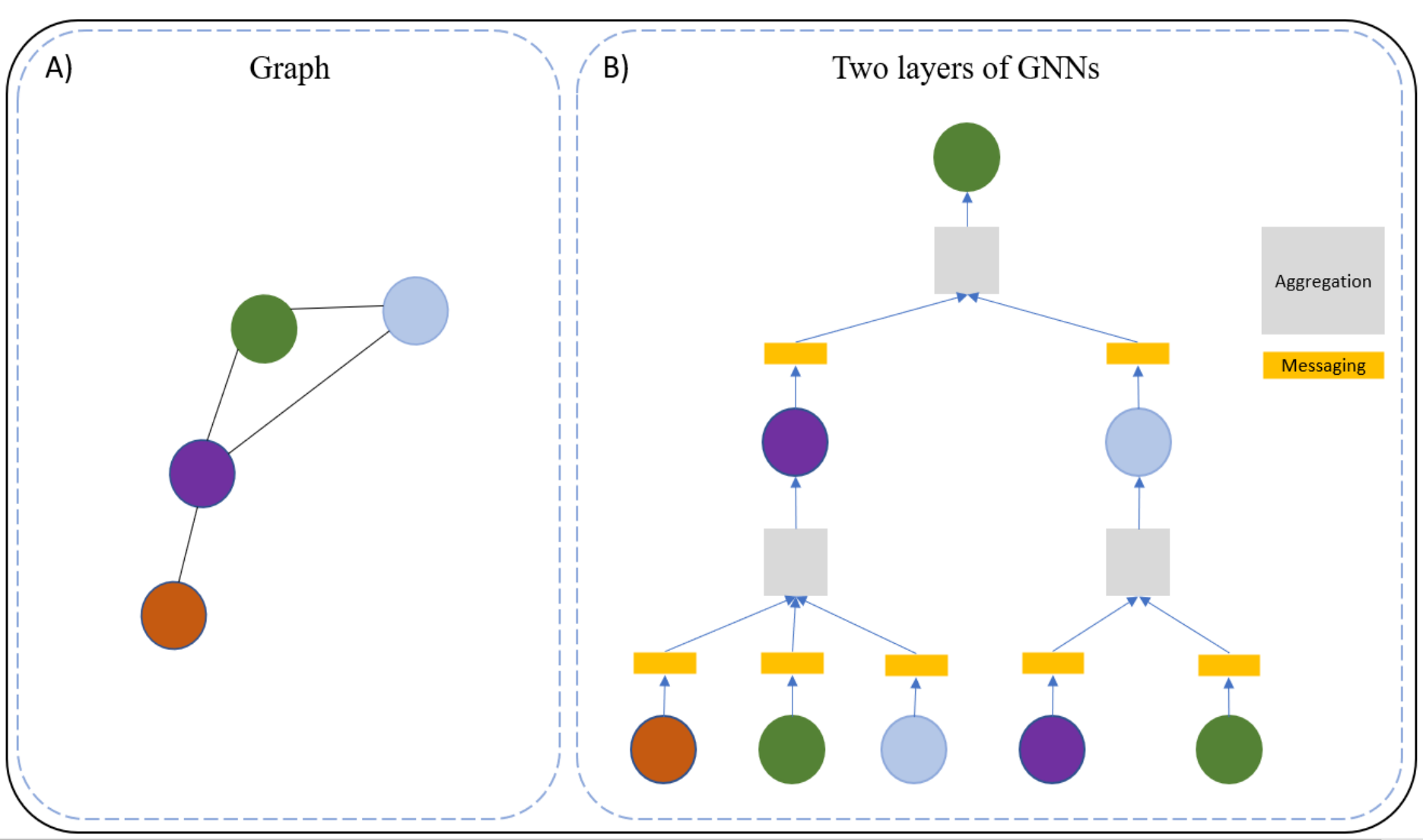

Figure 4 A) Graph with four nodes B) Corresponding GNN of the graph in part A, learning the representation of the green node based on the information gained from neighbors.

## Benchmark datasets:

There are several datasets used in the literature to evaluate the effectiveness of the DTI prediction models. Since study [7] has introduced almost all popular databases pertaining to this field, this section presents the four most common benchmark datasets: Yamanishi, Lou's, KIBA, and Davis.

### Yamanishi dataset:

This dataset, introduced in [25], has been widely utilized as a benchmark dataset. It is divided into four groups: enzymes, ion channels, G-protein-coupled receptors, and nuclear receptors. The dataset was collected from the KEGG BRITE [26], BRENDA [27], SuperTarget [28], and DrugBank [29] databases. **Error! Reference source not found.** illustrates some details of this dataset. Notably, some studies [30,31] have added new data to this database.

Table 2 Details of the Yamanishi dataset

| Group | No. drugs | No. proteins | No. interactions | Density |
|---|---|---|---|---|
| **Enzymes** | 445 | 664 | 2926 | 0.01 |
| **Ion channels** | 210 | 204 | 1476 | 0.034 |
| **GPCR** | 223 | 95 | 635 | 0.03 |
| **Nuclear receptors** | 54 | 26 | 90 | 0.064 |

### Luo's dataset:

Luo's dataset [32] consists of four node types and six edge types; drug nodes, drug-drug interactions, drug-target interactions, and protein nodes and protein-protein interactions have been extracted from DrugBank [33] and HPRD [34] databases, respectively. Disease nodes and their associations with drugs and proteins have been extracted from the Comparative Toxicogenomics Database (CTD) [35]. Finally, side effect nodes and their drug associations are obtained from the SIDER [36] database. This database has been mainly used when an approach involves heterogeneous networks and graph mining. The details of this dataset regarding nodes and edges are presented in and
Table 4, respectively.

Table 3 Details of nodes in Luo's dataset

| Node type | Number of nodes |
|---|---|
| **Drug** | 708 |
| **Protein** | 1512 |
| **Disease** | 5603 |
| **Side-effect** | 4192 |

Table 4 Details of edges in Luo's dataset

| Edge type | Number of edges | Density |
|---|---|---|
| **Drug-drug** | 10036 | 0.02 |
| **Drug-protein** | 1923 | 0.0018 |
| **Drug-disease** | 199214 | 0.05 |
| **Drug-side effect** | 80164 | 0.027 |
| **Protein-disease** | 1596745 | 0.188 |
| **Protein-protein** | 7363 | 0.003 |

### Davis and KIBA datasets:

Contrary to the above-mentioned datasets used for binary classification studies, the current two datasets are mainly used for drug-target binding affinity prediction, which is a regression task. The Davis [37] dataset uses dissociation constant ($K_d$) values to show the affinity between drugs and targets, whereas the KIBA [38] dataset adapts the KIBA value, a combination of three different sources, including $K_i$, $K_d$, and $IC_{50}$. The lower the value of either of these metrics, the stronger the binding affinity. Drugs and proteins with fewer than 10 interactions were filtered out from the KIBA dataset. The details of these datasets are shown in Table 5.

Table 5 Details of Davis and KIBA datasets

| Dataset | No. drugs | No. proteins | No. interactions | Density |
|---|---|---|---|---|
| **Davis** | 68 | 442 | 30056 | 1.0 |
| **KIBA** | 2111 | 229 | 118254 | 0.244 |

# Measurements:

Measurements based on the prediction task can be categorized into two groups: those used for classification and those utilized for regression analysis. There are several measurements being used for classification, including accuracy, specificity, precision, and recall, to name but a few. However, since DTI prediction datasets are significantly imbalanced, other metrics are more suitable than the aforementioned metrics, which are explained here.

## F1-score:

This metric provides a balanced measure of a model's precision and recall, ranging from zero to one. It is particularly useful when the class distribution of the datasets is imbalanced. The F1-score is calculated using Equation (5).

$$\mathrm{F1} = \frac{2\ \mathrm{Precision} \,.\, \mathrm{Recall}}{\mathrm{Precision} + \mathrm{Recall}} \tag{5}$$

## Matthews Correlation Coefficient (MCC):

This metric summarizes all four confusion matrix categories- true positive, true negative, false positive, and false negative- and is a more reliable statistical rate [39]. The formula for this metric is given by Equation **Error! Reference source not found.**:

$$\mathrm{MCC} = \frac{\mathrm{TP}.\mathrm{TN} - \mathrm{FP}.\mathrm{FN}}{\sqrt{(\mathrm{TP} + \mathrm{FP}) \,.\, (\mathrm{TP} + \mathrm{FN}) \,.\, (\mathrm{TN} + \mathrm{FP}) \,.\, (\mathrm{TN} + \mathrm{FN})}} \tag{6}$$

## AUROC and AUPR:

AUROC summarizes the ROC curve, a curve with TPR on the y-axis and FPR on the x-axis, into a single number, describing the capability of the model to distinguish between classes, whereas AUPR depicts the area under the precision-recall curve and is widely used when datasets are imbalanced. On these datasets, AUPR is a useful alternative to AUROC, highlighting the performance differences lost in AUROC [40].

## Mean square error (MSE):

MSE measures the average of the squared difference between actual values and the model's corresponding predicted value, indicating the closeness of the learned line to the true values. This metric, explained in Equation (*7*) is the most commonly used measurement in regression tasks.

$$\mathrm{MSE} = \frac{1}{n} \sum_{i=1}^{n} (p_i - y_i)^2 \tag{7}$$

## Concordance index (CI):

This metric indicates that considering two DTPs, the predicted binding affinity values are of the same order as their true values. It is calculated through Equations (8) and (9),

$$\mathrm{CI} = \frac{1}{Z} \sum_{a_i > a_j} h\left(p_i - p_j\right) \tag{8}$$

$$h(x) = \begin{cases} 1, & \text{if } x > 0 \\ 0.5, & \text{if } x = 0 \\ 0, & \text{if } x < 0 \end{cases} \tag{9}$$

where Z is the normalization constant, $a_i$ is the true affinity value, and $p_i$ is the predicted affinity value [41].

### Regression toward the mean (rm2):

This metric, calculated through Equation (*10*), has been employed to assess regression-based models, particularly quantitative structure-activity relationship (QSAR) models; it is a modified version of $r^2$ that indicates the square correlation between actual and predicted values. It also incorporates $r_0^2$, which is a squared correlation with zero intercepts [42].

$$rm^2 = r^2\left(1 - \sqrt{r^2 - r_0^2}\right) \tag{10}$$

## Network-based methods in DTI prediction:

In this section, we will discuss previous studies on DTI prediction. Although there have been some review papers [43–46] considering network-based methods, this paper includes a wider range of methods and approaches than those, including random walk-based, GNN-based, and metapath-based methods, as illustrated in Figure 5. It is also noteworthy to mention that there are several network-based methods [31,47–51] that are not based on these approaches and are not included in these methods. Also, different methods of constructing similarity networks are discussed prior to the main methods of the papers.

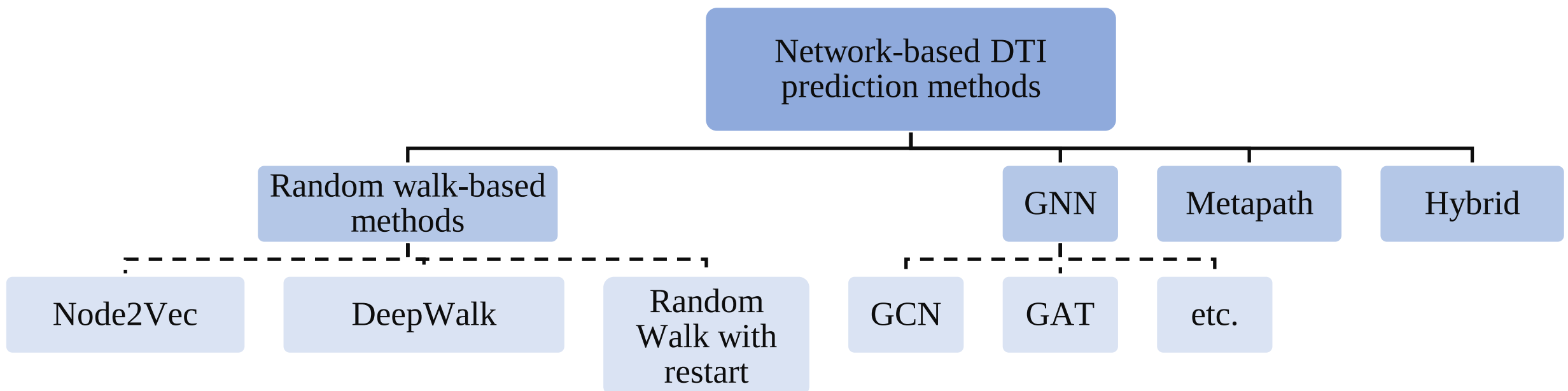

Figure 5 Examined categories of network-based methods in this paper.

### Constructing similarity networks:

On the one hand, some studies [52–55] leveraged the Jaccard similarity metric to construct two extra layers, drug-drug and protein-protein similarity layers, based on the existing layers of their network. These two networks are analyzed along with the other layers of a biological heterogeneous network (BHN). DTI-MGNN [56] utilizes this coefficient to convert each heterogeneous layer into a homogeneous one.

On the other hand, [56–60] adopted the Tanimoto coefficient [61] and Smith-waterman [62] score, which are based on the chemical structure of drugs and amino acid sequences of proteins, respectively, to build the similarity networks, where DTi2Vec [57] keeps the top k similarities of each node, as many of these links have a very low weight, incapable of providing useful information. In GCHN-DTI [59], drug-drug and protein-protein interaction networks have been linearly combined with similarity networks.

Affinity2Vec [15] builds two pairs of similarity networks; one is constructed based on the Tanimoto coefficient and Smith-Waterman score, while the other is generated based on the cosine similarity applied to the extracted features from the drug SMILES and the protein sequence. An autoencoder model has been utilized, incorporating multi-layer GRU networks for feature extraction [63]. Consequently, ProtVec [64] has been used to extract protein features. Although in NeoDTI [65], the Smith-Waterman score is used to construct the similarity network of the proteins, the corresponding network of drugs has been erected using dice similarities of the Morgan fingerprints [66], computed by RDKit [67].

In DTI-HETA [68], these networks are constructed using SNF [69], which is applied to the similarities of drug-side effects, drug chemical structure, drug physicochemical properties, co-pathway, protein-protein interaction network, and Gene Ontology.

BG-DTI [70] establishes similarity matrices considering both network and sequence similarities. Network similarity is calculated using the Jaccard similarity coefficient. In contrast, the sequence similarity of drugs is obtained based on the Tanimoto coefficient, and the sequence similarity of targets has been calculated by incorporating the Levenshtein similarity coefficient. GIAE-DTI [71] has calculated protein sequence similarity using Smith-Waterman, while others are the same as BG-DTI. In this research, like DTI-HETA and GCHN-DTI, the main networks have been dropped, and the fusion of similarity networks and the DTI network is analyzed.

## Random walk-based methods:

In [72], the cosine similarity measure has been applied to features learned by DeepWalk [73], followed by two inference methods, drug-based and target-based similarity inferences, to predict DTIs. It is shown that similarity computation using DeepWalk on the topology of a tripartite network (drug-disease-protein) outperforms chemical structure and genomic sequence-based similarities regarding AUC. Furthermore, their subsequent study [74] leverages node2vec, which uses a biased and more flexible random walk strategy, enabling it to provide both local and global samples to improve prediction performance. The authors also used a classification-based prediction method and the former inference-based methods to improve the results. DTi2Vec [57] and NEDTP [52], once they have learned a node's representation using node2vec, leverage XGBoost [75] to improve prediction performance. However, NEDTP categorizes negative samples into three groups: negative instances involving drugs and proteins, where both have positive samples, only one has positive samples, or neither has positive samples.

DTINet [32] and [76] attempted to tackle the issue of incomplete DTI datasets. In DTINet, a random walk with restart (RWR) has been employed on a BHN to obtain connectivity patterns between each drug and protein. To address the above-mentioned problem, the DCA method [77] has been adopted to obtain low-dimensional feature vectors. Then, it projects drug feature vectors (FVs) to protein space and further predicts the DTI. In [76], a weighted fusion of methods, including a drug similarity method based on the size of the common substructure named SIMCOMP [78], WNN-GIP [79], and random walk with restart, was applied to negative samples. Whenever the model predicts an interaction as positive, it will be added to the dataset, creating a revised dataset. Afterward, BLM-NII [80] is applied to the datasets, showing a superior prediction effect.

The generalized overview of the analyzed methods in this section is demonstrated in Table 6.

Table 6 Details of the random walk-based methods

| Study | Graph Mining Method | Dataset | Main Contribution | Source Code |
|---|---|---|---|---|
| [72] 2017 | DeepWalk | DrugBank, and Diseasome [81], Bio2rdf [82], UniProt [83], HGNC [84], OMIM [85] | Employing representation learned by DeepWalk outperforms chemical structures and genomic sequence-based similarities. | link |
| DTINet [32] 2017 | RWR | Luo's dataset | Incompleteness of the dataset is addressed. | link |
| [74] 2021 | node2vec | DrugBank, Bio2RDF, Diseasome, SIDER, KEGG [86], PharmGKB [86], PubChem [87], UniProt, HGNC, OMIM, UMLS [88], and MESH [89] | Adopting node2vec rather than DeepWalk has led to an improvement in prediction performance. | link |
| DTi2Vec [57] 2021 | node2vec | Yamanishi | The top k links of the similarity matrices augment the DTI network, reducing noise. | link |
| NEDTP [52] 2021 | node2vec | DrugBank, Therapeutic Target Database (TTD) [90], PharmGKB, ChEMBL [91], BindingDB [92], IUPHAR/BPS Guide to PHARMACOLOGY [93] | Negative samples are selected using a specific strategy. | link |
| [76] 2021 | RWR | Yamanishi | The same problem as DTINet is tackled with data augmentation. | - |

## GNN-based methods:

These methods can be divided into two categories: first, those methods that have applied GNNs to the graph representation of drugs or proteins, and second, those that have employed them in their main biological datasets. HampDTI [94] studied both drug molecule graphs and the biological network.

Adopting a 1D sequence representation results in losing significant structural information, so GraphDTA [95] converts drugs' SMILES to graph molecules using RDKit, where each node (atom) on the graph is represented using a set of atomic features adopted from DeepChem [96]. Then, to learn their embedding, four channels of GNNs, including GCN, GAT, GIN [97], and GCN-GAT, are applied separately. Nevertheless, proteins are represented by amino acid sequences, and three convolutional layers are used to learn their embeddings. DrugBAN [98] uses the same embedding process. Still, after that, instead of simply employing fully connected layers as a prediction module, a bilinear attention network learns local interaction representations by which the interpretability of the model is enhanced. In addition, the authors have attempted to address the cross-domain prediction problem by utilizing a conditional domain adversarial network (CDAN). This network transfers learned knowledge from the source domain to the target domain to improve cross-domain generalization.

GraphormerDTI [99] adopts 12 Graph Transformer layers on the drug molecular graph, while protein sequences are analyzed using 3 layers of CNN. The initial features of the atoms and amino acids are learnable parameters.

Furthermore, DGraphDTA [100] converts drug SMILES and protein sequences to their corresponding graph representation, followed by three GCN layers. To create protein graphs in this study, Pconsc4 [101], which is implemented using the U-net architecture, is utilized. The initial features of each drug and protein's node consist of 78 and 54 elements, respectively. In GEFormerDTA [102], GCN has been applied to drug molecules, resulting in a new graph, which is later concatenated with the protein graph. Another GCN has been used to reduce the dimensionality of this new graph. Meanwhile, this framework utilizes degree centrality encoding, atomic spatial position encoding, and interatomic chemical bonding coding as input to an encoder named ESC, consisting of attention and conv1d blocks. Furthermore, residual jumps have also been added to slow down the generalization performance of the network. GEFA [103] applies GCN on the drug molecules, followed by a max pooling layer, resulting in a single node. This node is then connected to the protein graph, and the weights of the edges between the drug node and the residue nodes indicate the likelihood of the contribution of the residues. To learn these weights, a self-attention mechanism is adopted.

The adjacency matrix of drugs and proteins is fixed in other studies; in GLCN-DTA [104], however, graph learning-convolutional network (GLCN) layers are employed to learn the soft adjacency matrix of protein and drug molecular graph. This module enhances the model's performance by optimizing and modifying graph topology, leading to learning better structural information regarding DTA prediction.

Despite other studies, KDBNet [105] uses the three-dimensional structure of proteins and drugs. To represent the 3D structure with graphs, the geometric feature of each atom or residue is added to the feature set of the nodes. Protein graphs in this study are analyzed using a Graph Transformer [106], while drug molecules are processed by geometric vector perceptrons (GVP) [107]. Since the 3D structure of all the proteins is not available, some proteins are eliminated from the datasets. KDBNet, in addition to the affinity prediction, estimates prediction uncertainties to avoid false positive results. Uncertainties are quantified by training eight independent model replicates. AttentionMGT-DTA [108] has employed 3D structure and 1D sequence representation of proteins. The former is obtained from AlphaFold2 [109] and is analyzed by Graph Transformer, while the latter is processed by ESM-2 [110]. These features are then integrated by a cross-attention module. Drugs are represented by graph molecules in this study.

The generalized overview of the analyzed GNN-based methods on drug molecules and protein sequences is shown in Table 7.

Table 7 Details of the GNN-based methods on the graph representation of drug molecules of protein sequences

| Study | Graph Mining Method | Dataset | Main Contribution | Source Code |
|---|---|---|---|---|
| GraphDTA [95] 2020 | GCN, GAT, GIN, and GCN-GAT | Davis and KIBA | A wide range of GNNs is employed on drug molecular graphs. | link |
| DGraphDTA [100] 2020 | GCN | Davis and KIBA | Both drugs and proteins are represented using graphs. | link |
| GEFA [103] | GCN | Davis | Early fusion of the drug | link |
| | | | node and protein graph leads to the detection of residues with a high likelihood of contributing to binding. | |
| DrugBAN [98] 2023 | GCN | BindingDB, BioSNAP [111], and Human [112] | The bilinear attention network has increased the interpretability, and CDAN has enhanced cross-domain prediction capability. | link |
| KDBNet [105] 2023 | Graph Transformer, GVP | Davis and KIBA | Utilization of the 3D structure of drugs and proteins and estimation of uncertainties. | link |
| GraphormerDTI [99] 2024 | Graph Transformer | DrugBank, Davis, and KIBA | A deep GNN is adopted to learn molecular features. | link |
| GEFormerDTA [102] 2024 | GCN | Davis and KIBA | Interpretability enhancement using self-attentive values among protein and drug nodes. | link |
| AttentionMGT-DTA [108] 2024 | Graph Transformer | Davis and KIBA | Adoption of AlphaFold2 for 3D structure prediction of proteins and utilization of a cross-attention module to integrate features from two modalities. | link |
| GLCN-DTA [104] 2024 | GLCN | Davis, KIBA, Metz [113], ToxCast [114] | Instead of using a fixed adjacency matrix, a soft one is utilized to improve the representation learning. | - |

Regarding the second group, NeoDTI [65] was proposed in 2018, which is mainly similar to GraphSAGE; however, since it is supposed to address heterogeneous networks, the method considers different sets of weights for each edge type. Also, the model is optimized to reconstruct the original weight matrix. EEG-DTI [53] leverages GCN, like NeoDTI, while overcoming the over-smoothing problem by concatenating the output of each layer. DTI-HETA [68] adopts GAT to highlight neighboring nodes' contributions and obtain more meaningful embeddings. Instead of a conventional message propagation method on a graph, PPAEDTI [58] introduces a PageRank-like method, enabling the model to consider both immediate and higher-order neighbors.

Nevertheless, GCHN-DTI [59] applies four GCN layers to learn graph embedding. To cope with the over-smoothing problem, the outputs of the GCN layers are fused using this mechanism. In BG-DTI [70], CNN blocks have been utilized to extract features from drug SMILES and protein sequences, which are then fed to GAT as initial features. To learn the BHN embedding, two GAT layers and a GCN layer are utilized. BG-DTI fuses the output of the first GAT with that of the GCN, the third layer, to overcome over-smoothing. In iGRLDTI [115], despite BG-DTI, an autoencoder is leveraged to learn a more compact representation of drugs obtained by RDKit from SMILES and a protein’s amino acids are grouped into four categories to reduce dimensionality. These are then fed to a GCN with NDLS [116], which is used to counteract the over-smoothing. Moreover, as a quantitative measure, Mean Absolute Distance (MAD) is applied to observe the over-smoothing of the representation being learned for nodes.

BITGNN [54] employs fingerprint and PsePSSM as feature representations for drugs and proteins. These features are then fed to four different GATs and GCNs based on the type of central node and neighbors, followed by a bidirectional transformer. HNetPa-DTI [117], however, adopts PubChem fingerprints, KSCTriad descriptors [118], and one-hot encoding vectors as initial features of drugs, proteins, and diseases, respectively. In this study, GraphSAGE is adopted to learn drug and protein representations based on the heterogeneous network. Meanwhile, another representation of proteins is learned by GCN on bipartite graphs of protein-GO terms and protein-pathways. Their main contribution is selecting high-quality negative samples. A negative sample is chosen if the shortest path length between the drug and protein on the drug-protein-disease network is higher than two and the shortest path between them on the drug-protein bipartite network equals three or five. Also, there must be at least one negative sample for every drug and protein among the positive samples.

In GIAE-DTI [71], the Weighted K nearest known neighbors (WKNKN) algorithm is applied to augment the DTI network. A graph autoencoder comprising a GIN-based encoder and two decoders, an edge decoder, and a degree decoder, is employed to learn the representation of drugs and proteins. The initial features of nodes are acquired by applying SVD decomposition on the matrix of $A = \begin{pmatrix} DD & Y_{DTI} \\ Y_{DTI}^T & TT \end{pmatrix}$. To increase the network connectivity, GENNDTI [119] introduces router nodes. Drugs with similar physicochemical properties and fractions of substructures are connected to the same router node, and for targets, similarities are calculated based on physicochemical properties, indices, and descriptors for amino acid sequences. To exploit drug-drug and target-target networks, GCN is adopted, while for the drug-target layer, an element-wise product performed better than GCN. Features extracted using these two modules, along with the Morgan fingerprint of drugs and composition of amino acid residues, dipeptides, and tripeptides of proteins, followed by PCA, are fed to GRU to fuse them.

In GCN-DTI [120], the constructed BHN is different from the ones mentioned above. In this study, a drug-protein pair (DPP) network is built; the combination of each drug and protein, in other words, creates a node in this graph, resulting in a massive graph with $D_n \times T_m$ nodes, where n and m are the numbers of drugs and proteins, and edges consist of three types, including strong and weak associations and non-associations with weights of 1, 0.5, and 0, respectively. Proteins are represented by a 26-dimensional feature consisting of 6 chemical characteristics and the relative proportions of each of the 20 amino acids, and drugs are represented by their chemical categories since they can illustrate the chemical properties of the drug. Then, a GCN network is used to integrate these features with the structural features of the DPP network. GVDTI [60] uses a wide variety of features. It adopts an attention mechanism at the attribute level on the BHN of drug-disease-protein, followed by a graph convolutional autoencoder (GCA) applied on similarity matrices, obtaining pairwise topological embedding vectors. A pairwise attribute-embedding matrix is built based on guilt by association, meaning drugs or proteins having interactions or associations with similar drugs, proteins, or diseases are more likely to interact. Finally, pairwise attribute distribution is learned by undertaking a convolutional variational autoencoder, which is applied to the adjacency matrix of the BHN.

Despite earlier models, GeNNius [121] emphasizes scalability and cross-dataset generalization. This model consists of four GraphSAGE layers followed by two dense layers. Its execution time is significantly lower than that of other state-of-the-art methods. Furthermore, the model exhibits high generalization capability for cross-dataset DTI inference, though it falls short when addressing imbalanced datasets.

The generalized overview of the analyzed GNN-based methods on a BHN is demonstrated in Table 8.

Table 8 Details of the GNN-based methods on a BHN

| **Study** | **Graph Mining Method** | **Dataset** | **Main Contribution** | **Source Code** |
|---|---|---|---|---|
| NeoDTI [65] 2018 | A GNN like GraphSAGE | Luo's dataset | Integration of neural message passing with topology-preserving optimization to extract latent drug and target features for DTI prediction. | link |
| EEG-DTI [53] 2021 | GCN | Yamanishi and Luo's datasets | The concatenation of the output of each layer has prevented over-smoothing. | link |
| GCN-DTI [120] 2021 | GCN | Yamanishi dataset and DrugBank | In the erected network, nodes indicate drug-target pairs. | link |
| PPAEDTI [58] 2022 | PPNP [122] | Yamanishi, HGBI [123], and the dataset proposed in [124] | An approximate personalized propagation layer is adopted instead of conventional message-passing layers. | - |
| DTI-HETA [68] 2022 | GCN and GAT | Yamanishi dataset and DrugBank | The contributions of neighboring nodes are | link |
| | | | highlighted using the GAT. | |
| GCHN-DTI [59] 2022 | GCN | Yamanishi and Luo's datasets | The attention mechanism is applied to the fuse output of each GCN layer. | - |
| GVDTI [60] 2022 | GCA | Luo's dataset | Pairwise topological embedding, attribute embedding, and attribute distribution are adopted. | - |
| BG-DTI [70]2023 | GCN and GAT | Yamanishi, Luo's, and CTD datasets | Similarity networks are built using both network and sequence-based metrics, and the initial features of GNNs are meaningful. | link |
| iGRLDTI [115] 2023 | GCN | Luo's dataset | NDLS has been adopted to overcome the problem of over-smoothing. | link |
| BiTGNN [54] 2024 | GCN, GAT | Yamanishi dataset | Bidirectional cross-attention influences drug and protein feature learning by considering protein and drug features, respectively. | - |
| HNetPa-DTI [117] 2024 | GraphSAGE, GCN | DrugBank, UniProt, QuickGO [125], KEGG, Reactome [126], WikiPathways [127], STRING [128], CTD | High-quality negative sample selection. | link |
| GIAE-DTI [71] 2024 | GIN | Luo's dataset | WKNKN algorithm is employed to augment the DTI network. | link |
| GENNDTI [119] 2024 | GCN | Davis and KIBA | The introduction of router nodes increases network connectivity. | link |
| GeNNius [121] 2024 | GraphSAGE | DrugBank, BindingDB, Davis, and Yamanishi datasets | The scalability and cross-dataset generalization of the proposed model surpasses other methods. | link |

## Metapath-based methods:

As different types of nodes do not share their feature space in heterogeneous networks, and GNNs are primarily introduced for homogeneous networks and are restricted to immediate neighbor aggregation [129], more metapath-based methods are being explored [130]; furthermore, metapaths construct high-order relationships between a starting node and an ending node, specifying the biological semantic relations in certain substructures, i.e., metagraphs [131]. Although GNNs can be repurposed for heterogeneous ones such as NeoDTI, they cause exponential growth in the number of model parameters.

In [132], BHN comprises 12 networks and 51 metapaths with a maximum length of four defined on this network. Features are embedded by multiplying adjacency matrices with respect to the order of the chosen metapath, which are then fed to a random forest and SVM to predict a DTI. Affinity2Vec [15] uses the same method as [132] to extract features from the BHN, while the number of predefined metapaths is six. Embedding-based features, explained in constructing similarity networks, are concatenated to these features and fed to XGBoost to make predictions. Since multimodal representation learning provides better generalization than a single source of information, DeepMPF [30] leverages multimodal representation learning, incorporating sequence, heterogeneous structure, and similarity modalities. Regarding sequence modality, features of drug SMILES and protein sequences are extracted using word2vec and a 3-mers sparse matrix, respectively. Six metapaths are defined to exploit heterogeneous structure information by maximizing the probability of each center node (word) using the metapath embedding model of CBOW. Finally, the similarity of structures has been exploited using Smith-Waterman scores and SIMCOMP. In MHTAN-DTI [55], seven metapaths are defined, and transformers are adopted to learn each metapath instance embedding regarding the central node and the metapath type. It has been shown that metapaths related to side effects have more significance than those related to diseases.
The generalized overview of the illustrated methods in this section is demonstrated in Table 9.

Table 9 Details of the metapath-based methods

| Study | Graph Mining Method | Dataset | Main Contribution | Source Code |
|---|---|---|---|---|
| [132] 2016 | 51 metapaths | Chem2Bio2RDF, DrugBank, and PubChem BioAssay database | A wide range of entities, including nine node types and 12 edge types, exists in the BHN. | - |
| Affinity2Vec [15] 2022 | six metapaths | Davis and KIBA | Construction of two pairs of drug and protein similarity networks and application of three feature extractors. | link |
| DeepMPF [30] 2023 | six metapaths | Yamanishi and proDB (DTIs are collected from DrugBank and [133]) | It includes three modalities: sequence, heterogeneous structure, and similarity. | link |
| MHTAN-DTI [55] 2023 | seven metapaths | Luo's dataset | Transformers are adopted for message passing between the nodes. | link |

## Hybrid methods:

Studies that have utilized a combination of methods from the groups above are explained in this section.
In SkipGNN [134], a skip graph is introduced in which second-order adjacent nodes are now directly connected. Also, the SkipGNN model is introduced, leveraging both the skip graph and the original graph for representation learning, uniquely suited for molecular interactions. Instead of simply concatenating the embedding of two graphs, the model fuses them through an iterative method. In this study, initial features are gained using node2vec.
In SGCL-DTI [135], five and four predefined metapaths are leveraged on drugs and proteins, respectively. Then, a GCN network is adopted to learn a representation of each node with an initial vector sampled from a standard normal distribution. Once these representations are learned, two DPP networks are constructed: semantic and topology graphs. Concerning the former, two nodes are connected if their cosine similarity of learned representation is among the top k, whereas two nodes are connected in the latter if they have a common drug or protein. After that, a supervised co-contrastive is adopted to predict interactions, and a contrastive objective is used to supervise the final task, prediction. One should notice that positive samples of each node of the topological network include the corresponding node in the semantic graph and its first-order neighbors with the same label. The exact process is applied to the semantic graph. DMHGNN [136] concatenates the adjacency matrix of each association and interaction to create an initial representation of drugs and proteins. For instance, drug-drug, drug-protein, and drug-disease are feature matrices of N drugs, and then, a denoising autoencoder (DAE) is applied for the dimension reduction of these vectors. These features, along with the features extracted by the first layer of GCNs, are used to create the semantic and topology graphs in the same way as SGCL-DTI. In addition to the previous graphs, a collaborative graph, which in fact is the mean of those graphs, is erected. Finally, the extracted features from these three graphs are fed to an attention layer to be integrated.
IMCHGAN [137] adopts GAT for each type of metagraph with random initial features, which are then integrated through the attention mechanism. In MHGNN [131], ten metapaths with a maximum length of five are defined. Along each metapath instance, all nodes' features, initialized as one-hot encoding, are considered. Then, features of drugs and proteins are updated with the corresponding GAT of that metapath. After integrating all features of drugs and proteins extracted from all metapath types, a graph of DTPs is constructed, and two GCN layers are applied for DTI prediction. AMGDTI [138] uses node2vec to learn the initial features of a GCN network like SkipGNN; however, it searches for an adaptive meta-graph from a heterogeneous network without requiring domain knowledge, enabling efficient integration of complex multiple semantic relationships and structure information embedded in the heterogeneous network. HampDTI [94] undertakes an adaptive selection of metapaths with some differences. AMGDTI applies some rules on selecting the following links randomly, while in HampDTI, a specific learnable weight is allocated for each link. HampDTI devises a node type-aware GNN based on SGC [139] to learn the representation of the constructed graph. The initial features of drugs are extracted from their molecular graphs, to which GCN is applied with initial features of 78-dimensional atomic attributions. In terms of targets, they are encoded into a 147-dimensional vector via Composition, Transition, and Distribution (CTD) rules.

In GADTI [140], prediction is conducted through a graph autoencoder model, in which the encoder consists of a GCN and RWR. This approach provides more information to nodes through a larger neighborhood while avoiding the over-smoothing and computational complexity caused by multi-layer convolutional networks. Here, DistMult [141] is employed as a decoder for link prediction. DTIHNC [142] applies a DAE in the same way as DMHGNN to extract features from the multiple layers of BHN. These features are then fed to GAT as initial features. Similar to GADTI, a GNN (a GAT here) and RWR are leveraged to learn the representation of BHN. These embeddings and features learned using the Jaccard similarity coefficient, named cross-modal similarities, are fed to the prediction module. DTI-MGNN [56] adopts RWR on constructed networks, gaining the probability distribution of all nodes, followed by a DAE on the concatenation of these matrices to obtain a low-dimensional feature representation. Similar to the SGCL-DTI, two graphs are constructed: one based on topology and the other on features. Afterward, two parallel GATs are applied to these networks, while a GCN is adopted with shared weights on both graphs to learn common features, enhancing the ability to fuse topological structures and semantic features simultaneously.

The generalized overview of the analyzed methods in this section is demonstrated in Table 10.

Table 10 Details of hybrid methods

| Study | Graph Mining Method | Dataset | Main Contribution | Source Code |
|---|---|---|---|---|
| SkipGNN [134] 2020 | Node2vec and SkipGNN | BioSNAP | This model is uniquely suited for molecular interactions. | link |
| GADTI [140] 2021 | GCN, RWR, and DistMult | Luo’s dataset | Instead of stacking GNNs, RWR is adopted to reach further neighbors and address the over-smoothing problem. | link |
| DTI-MGNN [56] 2021 | RWR, GAT, and GCN | Their dataset, based on Luo’s dataset | The capability of fusing topological structures and semantic features is enhanced. | link |
| SGCL-DTI [135] 2022 | Nine metapaths, GCN | Luo’s, Zheng's [143], and Yamanishi datasets | A new positive sample selection strategy is proposed in terms of contrastive learning. | link |
| IMCHGAN [137] 2022 | ten metapaths, GAT, and GCN | Luo’s dataset and datasets | The features learned using different metapaths and GATs are aggregated using an attention mechanism. | link |
| HampDTI [94] 2022 | GCN, adaptive meta-graph, SGC-based GNN | Luo’s dataset | The initial features of GNNs are meaningful and from different modalities, and metapaths are extracted automatically. | link |
| DTIHNC [142] 2022 | GAT, RWR, and Jaccard similarity coefficient | Luo’s dataset | Initial features of the GAT are extracted from adjacency matrices, and the cross-modal similarities are calculated by the Jaccard similarity coefficient. | - |
| MHGNN [131] 2023 | Ten metapaths, GAT | Luo’s, its extended version, and Zheng's datasets | Features of all nodes along each metapath are considered. | link |
| AMGDTI [138] 2024 | Adaptive meta- | Luo’s and Zheng's datasets | The method requires no | link |
| | graph, node2vec, GCN | | domain knowledge to extract metapaths and metagraphs. | |
| DMHGNN [136] 2025 | Meta-path guided GCN | Extended version of the Luo's dataset | The extended version of Luo's dataset is gathered. | link |

## Discussion and future challenges

Future challenges in this field can be divided into data-related and method-related challenges. The former includes the challenges associated with datasets, such as their size, diversity, and imbalance, whereas the latter is related to the approaches, including the generalization capability and interpretability of the methods, their evaluation, and the representation of data.

**Data-related challenges:** As discussed earlier, the prediction of the binding affinity between drugs and proteins is a more meaningful task, requiring a comprehensive benchmark dataset. The most commonly used datasets for regression are the KIBA and Davis datasets. Still, they do not incorporate additional modalities, such as drug-disease, drug-drug, and protein-disease associations, to name but a few. Hence, constructing a dataset including these diverse interactions and associations should be one of the key research priorities. Considering the classification task, the BETA dataset introduced in [144], including 0.97 million biomedical nodes and 8.5 million associations, is being overlooked. This benchmark evaluates DTI predictors across seven validation tasks. Moreover, one should notice that these datasets do not include negative samples, and the random selection of unknown data could be misleading. Thus, future studies should have a strategy for a negative sample selection, like HNetPa-DTI [117].

**Method-related challenges:** Considering current benchmark datasets, future studies should concentrate on models' generalization capability and interpretability, as proposed ones in recent studies have achieved a decent performance, e.g., MHGNN and AMGDTI have AUC and AUPR of over 97% on Lou's dataset.

- **Generalization:** Achieving higher generalization is possible using multimodality. For instance, these models can take advantage of the chemical structure of drugs and the amino acid sequence of proteins, along with information offered by biological networks like HampDTI. Many studies have utilized random or one-hot encoded vectors as initial features for GNNs, which can be replaced with features related to the discussed characteristics of drugs and proteins. SkipGNN and AMGDTI have employed features extracted by node2vec as initial inputs, considering the structural features of networks similar to GNNs. Although node2vec can reach further neighbors, is more global than GNNs, and enhances model performance, it still primarily focuses on networks.

  To evaluate the generalization capability of methods, as discussed in [144], studies should focus on inductive models. Study [145] illustrates cold-drug and cold-target settings, while cross-dataset evaluation results are generally more reliable as they help prevent any form of data leakage. For instance, regarding Luo's dataset, the extended version of this dataset is introduced in [46] and [135], so future studies can demonstrate their generalization by training their model on Luo's dataset and testing it on the extended version.
- **Interpretability and Explainability:** Some research, such as DrugBAN and GEFormerDTA, has considered the interpretability of their models, whereas there are more expansive areas that need to be explored. A case in point is which part of a drug molecule and protein sequence has higher significance in their interaction. Moreover, models like GNAN [146], which can be visualized, enabling humans to understand the flow and reasoning process, have been introduced recently and can fill the gap of interpretability in this field.
- **Evaluation:** Regarding the evaluation of classification models, accuracy is the most popular metric among studies [147]. However, as discussed earlier, data in this field are heavily imbalanced, so metrics such as AUPR and F1-score are more suitable. Furthermore, while reporting the AUPR metric, we strongly recommend that precision and recall be reported as they have different meanings and should not be overlooked. Another critical point is that some studies sample negative data at a 1:1 ratio, while this imbalance between classes exists in real-world scenarios. Therefore, it should be demonstrated that models are robust against this problem.

  Additionally, many of the previous studies are limited to point prediction, whereas from a practical standpoint, it is recommended that models should provide uncertainty estimation. As these predictions require validation in wet lab experiments, reporting the confidence of the predictions is desired. Conformal prediction [148] is one method that can be adopted for uncertainty quantification.

- **Three-dimensional representation:** The lack of 3D structure in proteins inhibits the adoption of methods that take advantage of this representation. This barrier is nearly overcome with the emergence of models predicting protein 3D structures with high accuracy. In particular, AlphaFold3 [149] can be employed to improve the accuracy and generalization of the methods.

# References

[1] S. Sinha, D. Vohora, Chapter 2 - Drug Discovery and Development: An Overview, (2018) 19–32. https://doi.org/https://doi.org/10.1016/B978-0-12-802103-3.00002-X.

[2] S. Kraljevic, P.J. Stambrook, K. Pavelic, Accelerating drug discovery, EMBO Rep. 5 (2004) 837–842. https://doi.org/https://doi.org/10.1038/sj.embor.7400236.

[3] R. Chen, X. Liu, S. Jin, J. Lin, J. Liu, Machine Learning for Drug-Target Interaction Prediction, Molecules 23 (2018). https://doi.org/10.3390/molecules23092208.

[4] S. Khot, S. Naykude, P. Adnaik, An overview of drug drug development process: Short Communication, J. Pharma Insights Res. 1 (2023) 67–74. https://jopir.in/index.php/journals/article/view/37.

[5] T.T. Ashburn, K.B. Thor, Drug repositioning: identifying and developing new uses for existing drugs, Nat. Rev. Drug Discov. 3 (2004) 673–683. https://doi.org/10.1038/nrd1468.

[6] C. Abels, M. Soeberdt, Can we teach old drugs new tricks?—Repurposing of neuropharmacological drugs for inflammatory skin diseases, Exp. Dermatol. 28 (2019) 1002–1009. https://doi.org/https://doi.org/10.1111/exd.13987.

[7] M. Bagherian, E. Sabeti, K. Wang, M.A. Sartor, Z. Nikolovska-Coleska, K. Najarian, Machine learning approaches and databases for prediction of drug–target interaction: a survey paper, Brief. Bioinform. 22 (2021) 247–269. https://doi.org/10.1093/bib/bbz157.

[8] G. Sotiropoulou, E. Zingkou, G. Pampalakis, Redirecting drug repositioning to discover innovative cosmeceuticals, Exp. Dermatol. 30 (2021) 628–644. https://doi.org/https://doi.org/10.1111/exd.14299.

[9] D. Sardana, C. Zhu, M. Zhang, R.C. Gudivada, L. Yang, A.G. Jegga, Drug repositioning for orphan diseases, Brief. Bioinform. 12 (2011) 346–356. https://doi.org/10.1093/bib/bbr021.

[10] Z. Xia, L.-Y. Wu, X. Zhou, S.T.C. Wong, Semi-supervised drug-protein interaction prediction from heterogeneous biological spaces, BMC Syst. Biol. 4 (2010) S6. https://doi.org/10.1186/1752-0509-4-S2-S6.

[11] L. Hood, L. Rowen, The Human Genome Project: big science transforms biology and medicine, Genome Med. 5 (2013) 79. https://doi.org/10.1186/gm483.

[12] B. Lomenick, R.W. Olsen, J. Huang, Identification of Direct Protein Targets of Small Molecules, ACS Chem. Biol. 6 (2011) 34–46. https://doi.org/10.1021/cb100294v.

[13] S.K. Middha, T. Usha, S. Sukhralia, C. Pareek, R. Yadav, R. Agnihotri, J. Tasneem, A.K. Goyal, D. Babu, Chapter 23 - Prediction of drug–target interaction —a helping hand in drug repurposing, in: A. Parihar, R. Khan, A. Kumar, A.K. Kaushik, H.B.T.-C.A. for N.T. and D.D. to M.S.-C.-2 I. Gohel (Eds.), Academic Press, 2022: pp. 519–536. https://doi.org/https://doi.org/10.1016/B978-0-323-91172-6.00006-6.

[14] T. Pahikkala, A. Airola, S. Pietilä, S. Shakyawar, A. Szwajda, J. Tang, T. Aittokallio, Toward more realistic drug–target interaction predictions, Brief. Bioinform. 16 (2015) 325–337. https://doi.org/10.1093/bib/bbu010.

[15] M.A. Thafar, M. Alshahrani, S. Albaradei, T. Gojobori, M. Essack, X. Gao, Affinity2Vec: drug-target binding affinity prediction through representation learning, graph mining, and machine learning, Sci. Rep. 12 (2022) 4751. https://doi.org/10.1038/s41598-022-08787-9.

[16] T. Mikolov, K. Chen, G. Corrado, J. Dean, Efficient estimation of word representations in vector space, ArXiv Prepr. ArXiv1301.3781 (2013).

[17] C. Knox, M. Wilson, C.M. Klinger, M. Franklin, E. Oler, A. Wilson, A. Pon, J. Cox, N.E. (Lucy) Chin, S.A. Strawbridge, M. Garcia-Patino, R. Kruger, A. Sivakumaran, S. Sanford, R. Doshi, N. Khetarpal, O. Fatokun, D. Doucet, A. Zubkowski, D.Y. Rayat, H. Jackson, K. Harford, A. Anjum, M. Zakir, F. Wang, S. Tian, B. Lee, J. Liigand, H. Peters, R.Q. (Rachel) Wang, T. Nguyen, D. So, M. Sharp, R. da Silva, C. Gabriel, J. Scantlebury, M. Jasinski, D. Ackerman, T. Jewison, T. Sajed, V. Gautam, D.S. Wishart, DrugBank 6.0: the DrugBank Knowledgebase for 2024, Nucleic Acids Res. 52 (2024) D1265–D1275. https://doi.org/10.1093/nar/gkad976.

[18] R. Interdonato, M. Atzmueller, S. Gaito, R. Kanawati, C. Largeron, A. Sala, Feature-rich networks: going beyond complex network topologies, Appl. Netw. Sci. 4 (2019) 4. https://doi.org/10.1007/s41109-019-0111-x.

[19] Y. Dong, N. V Chawla, A. Swami, metapath2vec: Scalable Representation Learning for Heterogeneous

Networks, in: Proc. 23rd ACM SIGKDD Int. Conf. Knowl. Discov. Data Min., Association for Computing Machinery, New York, NY, USA, 2017: pp. 135–144. https://doi.org/10.1145/3097983.3098036.
[20] A. Grover, J. Leskovec, node2vec: Scalable Feature Learning for Networks, in: Proc. 22nd ACM SIGKDD Int. Conf. Knowl. Discov. Data Min., Association for Computing Machinery, New York, NY, USA, 2016: pp. 855–864. https://doi.org/10.1145/2939672.2939754.
[21] J. Wang, A survey on graph neural networks, EAI Endorsed Trans. e-Learning 8 (2023) e6. https://doi.org/10.4108/eetel.3466.
[22] T.N. Kipf, M. Welling, Semi-Supervised Classification with Graph Convolutional Networks, in: Int. Conf. Learn. Represent., 2017.
[23] W. Hamilton, Z. Ying, J. Leskovec, Inductive Representation Learning on Large Graphs, in: I. Guyon, U. Von Luxburg, S. Bengio, H. Wallach, R. Fergus, S. Vishwanathan, R. Garnett (Eds.), Adv. Neural Inf. Process. Syst., Curran Associates, Inc., 2017. https://proceedings.neurips.cc/paper_files/paper/2017/file/5dd9db5e033da9c6fb5ba83c7a7ebea9-Paper.pdf.
[24] P. Veličković, G. Cucurull, A. Casanova, A. Romero, P. Liò, Y. Bengio, Graph Attention Networks, in: Int. Conf. Learn. Represent., 2018. https://openreview.net/forum?id=rJXMpikCZ.
[25] Y. Yamanishi, M. Araki, A. Gutteridge, W. Honda, M. Kanehisa, Prediction of drug–target interaction networks from the integration of chemical and genomic spaces, Bioinformatics 24 (2008) i232–i240. https://doi.org/10.1093/bioinformatics/btn162.
[26] M. Kanehisa, S. Goto, M. Hattori, K.F. Aoki-Kinoshita, M. Itoh, S. Kawashima, T. Katayama, M. Araki, M. Hirakawa, From genomics to chemical genomics: new developments in KEGG, Nucleic Acids Res. 34 (2006) D354–D357. https://doi.org/10.1093/nar/gkj102.
[27] I. Schomburg, A. Chang, C. Ebeling, M. Gremse, C. Heldt, G. Huhn, D. Schomburg, BRENDA, the enzyme database: updates and major new developments, Nucleic Acids Res. 32 (2004) D431–D433. https://doi.org/10.1093/nar/gkh081.
[28] S. Günther, M. Kuhn, M. Dunkel, M. Campillos, C. Senger, E. Petsalaki, J. Ahmed, E.G. Urdiales, A. Gewiess, L.J. Jensen, R. Schneider, R. Skoblo, R.B. Russell, P.E. Bourne, P. Bork, R. Preissner, SuperTarget and Matador: resources for exploring drug-target relationships, Nucleic Acids Res. 36 (2008) D919–D922. https://doi.org/10.1093/nar/gkm862.
[29] D.S. Wishart, C. Knox, A.C. Guo, D. Cheng, S. Shrivastava, D. Tzur, B. Gautam, M. Hassanali, DrugBank: a knowledgebase for drugs, drug actions and drug targets, Nucleic Acids Res. 36 (2008) D901–D906. https://doi.org/10.1093/nar/gkm958.
[30] Z.-H. Ren, Z.-H. You, Q. Zou, C.-Q. Yu, Y.-F. Ma, Y.-J. Guan, H.-R. You, X.-F. Wang, J. Pan, DeepMPF: deep learning framework for predicting drug–target interactions based on multi-modal representation with meta-path semantic analysis, J. Transl. Med. 21 (2023) 48. https://doi.org/10.1186/s12967-023-03876-3.
[31] Y. Chu, X. Shan, T. Chen, M. Jiang, Y. Wang, Q. Wang, D.R. Salahub, Y. Xiong, D.-Q. Wei, DTI-MLCD: predicting drug-target interactions using multi-label learning with community detection method, Brief. Bioinform. 22 (2021) bbaa205. https://doi.org/10.1093/bib/bbaa205.
[32] Y. Luo, X. Zhao, J. Zhou, J. Yang, Y. Zhang, W. Kuang, J. Peng, L. Chen, J. Zeng, A network integration approach for drug-target interaction prediction and computational drug repositioning from heterogeneous information, Nat. Commun. 8 (2017) 573. https://doi.org/10.1038/s41467-017-00680-8.
[33] C. Knox, V. Law, T. Jewison, P. Liu, S. Ly, A. Frolkis, A. Pon, K. Banco, C. Mak, V. Neveu, Y. Djoumbou, R. Eisner, A.C. Guo, D.S. Wishart, DrugBank 3.0: a comprehensive resource for 'Omics' research on drugs, Nucleic Acids Res. 39 (2011) D1035–D1041. https://doi.org/10.1093/nar/gkq1126.
[34] T.S. Keshava Prasad, R. Goel, K. Kandasamy, S. Keerthikumar, S. Kumar, S. Mathivanan, D. Telikicherla, R. Raju, B. Shafreen, A. Venugopal, L. Balakrishnan, A. Marimuthu, S. Banerjee, D.S. Somanathan, A. Sebastian, S. Rani, S. Ray, C.J. Harrys Kishore, S. Kanth, M. Ahmed, M.K. Kashyap, R. Mohmood, Y.L. Ramachandra, V. Krishna, B.A. Rahiman, S. Mohan, P. Ranganathan, S. Ramabadran, R. Chaerkady, A. Pandey, Human Protein Reference Database—2009 update, Nucleic Acids Res. 37 (2009) D767–D772. https://doi.org/10.1093/nar/gkn892.
[35] A.P. Davis, C.G. Murphy, R. Johnson, J.M. Lay, K. Lennon-Hopkins, C. Saraceni-Richards, D. Sciaky, B.L. King, M.C. Rosenstein, T.C. Wiegers, C.J. Mattingly, The Comparative Toxicogenomics Database: update 2013, Nucleic Acids Res. 41 (2013) D1104–D1114. https://doi.org/10.1093/nar/gks994.
[36] M. Kuhn, M. Campillos, I. Letunic, L.J. Jensen, P. Bork, A side effect resource to capture phenotypic effects of drugs, Mol. Syst. Biol. 6 (2010) 343. https://doi.org/https://doi.org/10.1038/msb.2009.98.
[37] M.I. Davis, J.P. Hunt, S. Herrgard, P. Ciceri, L.M. Wodicka, G. Pallares, M. Hocker, D.K. Treiber, P.P. Zarrinkar, Comprehensive analysis of kinase inhibitor selectivity, Nat. Biotechnol. 29 (2011) 1046–1051.

https://doi.org/10.1038/nbt.1990.
[38] J. Tang, A. Szwajda, S. Shakyawar, T. Xu, P. Hintsanen, K. Wennerberg, T. Aittokallio, Making Sense of Large-Scale Kinase Inhibitor Bioactivity Data Sets: A Comparative and Integrative Analysis, J. Chem. Inf. Model. 54 (2014) 735–743. https://doi.org/10.1021/ci400709d.
[39] D. Chicco, G. Jurman, The advantages of the Matthews correlation coefficient (MCC) over F1 score and accuracy in binary classification evaluation, BMC Genomics 21 (2020) 6. https://doi.org/10.1186/s12864-019-6413-7.
[40] K. Boyd, K.H. Eng, C.D. Page, Area under the Precision-Recall Curve: Point Estimates and Confidence Intervals BT - Machine Learning and Knowledge Discovery in Databases, in: H. Blockeel, K. Kersting, S. Nijssen, F. Železný (Eds.), Springer Berlin Heidelberg, Berlin, Heidelberg, 2013: pp. 451–466.
[41] M. Gönen, G. Heller, Concordance probability and discriminatory power in proportional hazards regression, Biometrika 92 (2005) 965–970. https://doi.org/10.1093/biomet/92.4.965.
[42] K. Roy, P. Chakraborty, I. Mitra, P.K. Ojha, S. Kar, R.N. Das, Some case studies on application of "r2" metrics for judging quality of quantitative structure–activity relationship predictions: Emphasis on scaling of response data, J. Comput. Chem. 34 (2013) 1071–1082. https://doi.org/https://doi.org/10.1002/jcc.23231.
[43] F. Cheng, C. Liu, J. Jiang, W. Lu, W. Li, G. Liu, W. Zhou, J. Huang, Y. Tang, Prediction of Drug-Target Interactions and Drug Repositioning via Network-Based Inference, PLOS Comput. Biol. 8 (2012) e1002503. https://doi.org/10.1371/journal.pcbi.1002503.
[44] Z. Zhang, L. Chen, F. Zhong, D. Wang, J. Jiang, S. Zhang, H. Jiang, M. Zheng, X. Li, Graph neural network approaches for drug-target interactions, Curr. Opin. Struct. Biol. 73 (2022) 102327. https://doi.org/https://doi.org/10.1016/j.sbi.2021.102327.
[45] Z. Wu, W. Li, G. Liu, Y. Tang, Network-Based Methods for Prediction of Drug-Target Interactions, Front. Pharmacol. Volume 9- (2018). https://www.frontiersin.org/journals/pharmacology/articles/10.3389/fphar.2018.01134.
[46] Y. Zhang, Y. Hu, N. Han, A. Yang, X. Liu, H. Cai, A survey of drug-target interaction and affinity prediction methods via graph neural networks, Comput. Biol. Med. 163 (2023) 107136. https://doi.org/https://doi.org/10.1016/j.compbiomed.2023.107136.
[47] X. Zeng, S. Zhu, W. Lu, Z. Liu, J. Huang, Y. Zhou, J. Fang, Y. Huang, H. Guo, L. Li, B.D. Trapp, R. Nussinov, C. Eng, J. Loscalzo, F. Cheng, Target identification among known drugs by deep learning from heterogeneous networks, Chem. Sci. 11 (2020) 1775–1797. https://doi.org/10.1039/C9SC04336E.
[48] X. Zeng, S. Zhu, Y. Hou, P. Zhang, L. Li, J. Li, L.F. Huang, S.J. Lewis, R. Nussinov, F. Cheng, Network-based prediction of drug–target interactions using an arbitrary-order proximity embedded deep forest, Bioinformatics 36 (2020) 2805–2812. https://doi.org/10.1093/bioinformatics/btaa010.
[49] Q. Ye, C.-Y. Hsieh, Z. Yang, Y. Kang, J. Chen, D. Cao, S. He, T. Hou, A unified drug–target interaction prediction framework based on knowledge graph and recommendation system, Nat. Commun. 12 (2021) 6775. https://doi.org/10.1038/s41467-021-27137-3.
[50] D. Zhou, Z. Xu, W. Li, X. Xie, S. Peng, MultiDTI: drug–target interaction prediction based on multi-modal representation learning to bridge the gap between new chemical entities and known heterogeneous network, Bioinformatics 37 (2021) 4485–4492. https://doi.org/10.1093/bioinformatics/btab473.
[51] X. Su, P. Hu, H. Yi, Z. You, L. Hu, Predicting Drug-Target Interactions Over Heterogeneous Information Network, IEEE J. Biomed. Heal. Informatics 27 (2023) 562–572. https://doi.org/10.1109/JBHI.2022.3219213.
[52] Q. An, L. Yu, A heterogeneous network embedding framework for predicting similarity-based drug-target interactions, Brief. Bioinform. 22 (2021) bbab275. https://doi.org/10.1093/bib/bbab275.
[53] J. Peng, Y. Wang, J. Guan, J. Li, R. Han, J. Hao, Z. Wei, X. Shang, An end-to-end heterogeneous graph representation learning-based framework for drug–target interaction prediction, Brief. Bioinform. 22 (2021) bbaa430. https://doi.org/10.1093/bib/bbaa430.
[54] Q. Zhang, C. He, X. Qin, P. Yang, J. Kong, Y. Mao, D. Li, BiTGNN: Prediction of drug–target interactions based on bidirectional transformer and graph neural network on heterogeneous graph, Int. J. Biomath. (2024). https://doi.org/10.1142/S1793524524500256.
[55] R. Zhang, Z. Wang, X. Wang, Z. Meng, W. Cui, MHTAN-DTI: Metapath-based hierarchical transformer and attention network for drug–target interaction prediction, Brief. Bioinform. 24 (2023) bbad079. https://doi.org/10.1093/bib/bbad079.
[56] Y. Li, G. Qiao, K. Wang, G. Wang, Drug–target interaction predication via multi-channel graph neural networks, Brief. Bioinform. 23 (2022) bbab346. https://doi.org/10.1093/bib/bbab346.
[57] M.A. Thafar, R.S. Olayan, S. Albaradei, V.B. Bajic, T. Gojobori, M. Essack, X. Gao, DTi2Vec: Drug–target

interaction prediction using network embedding and ensemble learning, J. Cheminform. 13 (2021) 71. https://doi.org/10.1186/s13321-021-00552-w.
[58] Y.-C. Li, Z.-H. You, C.-Q. Yu, L. Wang, L. Wong, L. Hu, P.-W. Hu, Y.-A. Huang, PPAEDTI: Personalized Propagation Auto-Encoder Model for Predicting Drug-Target Interactions, IEEE J. Biomed. Heal. Informatics 27 (2023) 573–582. https://doi.org/10.1109/JBHI.2022.3217433.
[59] W. Wang, S. Liang, M. Yu, D. Liu, H. Zhang, X. Wang, Y. Zhou, GCHN-DTI: Predicting drug-target interactions by graph convolution on heterogeneous networks, Methods 206 (2022) 101–107. https://doi.org/https://doi.org/10.1016/j.ymeth.2022.08.016.
[60] P. Xuan, M. Fan, H. Cui, T. Zhang, T. Nakaguchi, GVDTI: graph convolutional and variational autoencoders with attribute-level attention for drug–protein interaction prediction, Brief. Bioinform. 23 (2022) bbab453. https://doi.org/10.1093/bib/bbab453.
[61] P. Willett, Similarity-based virtual screening using 2D fingerprints, Drug Discov. Today 11 (2006) 1046–1053. https://doi.org/https://doi.org/10.1016/j.drudis.2006.10.005.
[62] T.F. Smith, M.S. Waterman, Identification of common molecular subsequences, J. Mol. Biol. 147 (1981) 195–197. https://doi.org/https://doi.org/10.1016/0022-2836(81)90087-5.
[63] M. Jang, S. Seo, P. Kang, Recurrent neural network-based semantic variational autoencoder for Sequence-to-sequence learning, Inf. Sci. (Ny). 490 (2019) 59–73. https://doi.org/https://doi.org/10.1016/j.ins.2019.03.066.
[64] E. Asgari, M.R.K. Mofrad, Continuous Distributed Representation of Biological Sequences for Deep Proteomics and Genomics, PLoS One 10 (2015) e0141287. https://doi.org/10.1371/journal.pone.0141287.
[65] F. Wan, L. Hong, A. Xiao, T. Jiang, J. Zeng, NeoDTI: neural integration of neighbor information from a heterogeneous network for discovering new drug–target interactions, Bioinformatics 35 (2019) 104–111. https://doi.org/10.1093/bioinformatics/bty543.
[66] D. Rogers, M. Hahn, Extended-Connectivity Fingerprints, J. Chem. Inf. Model. 50 (2010) 742–754. https://doi.org/10.1021/ci100050t.
[67] G. Landrum, Rdkit documentation, (n.d.).
[68] K. Shao, Y. Zhang, Y. Wen, Z. Zhang, S. He, X. Bo, DTI-HETA: prediction of drug–target interactions based on GCN and GAT on heterogeneous graph, Brief. Bioinform. 23 (2022) bbac109. https://doi.org/10.1093/bib/bbac109.
[69] B. Wang, A.M. Mezlini, F. Demir, M. Fiume, Z. Tu, M. Brudno, B. Haibe-Kains, A. Goldenberg, Similarity network fusion for aggregating data types on a genomic scale., Nat. Methods 11 (2014) 333–337. https://doi.org/10.1038/nmeth.2810.
[70] L. Liu, Q. Zhang, Y. Wei, Q. Zhao, B. Liao, A Biological Feature and Heterogeneous Network Representation Learning-Based Framework for Drug–Target Interaction Prediction, Molecules 28 (2023). https://doi.org/10.3390/molecules28186546.
[71] M. Wang, X. Lei, L. Liu, J. Chen, F.-X. Wu, GIAE-DTI: Predicting Drug-Target Interactions Based on Heterogeneous Network and GIN-based Graph Autoencoder, IEEE J. Biomed. Heal. Informatics (2024) 1–14. https://doi.org/10.1109/JBHI.2024.3458794.
[72] N. Zong, H. Kim, V. Ngo, O. Harismendy, Deep mining heterogeneous networks of biomedical linked data to predict novel drug–target associations, Bioinformatics 33 (2017) 2337–2344. https://doi.org/10.1093/bioinformatics/btx160.
[73] B. Perozzi, R. Al-Rfou, S. Skiena, DeepWalk: online learning of social representations, in: Proc. 20th ACM SIGKDD Int. Conf. Knowl. Discov. Data Min., Association for Computing Machinery, New York, NY, USA, 2014: pp. 701–710. https://doi.org/10.1145/2623330.2623732.
[74] N. Zong, R.S.N. Wong, Y. Yu, A. Wen, M. Huang, N. Li, Drug–target prediction utilizing heterogeneous bio-linked network embeddings, Brief. Bioinform. 22 (2021) 568–580. https://doi.org/10.1093/bib/bbz147.
[75] J.H. Friedman, Greedy Function Approximation: A Gradient Boosting Machine, Ann. Stat. 29 (2001) 1189–1232. http://www.jstor.org/stable/2699986.
[76] S. Liu, J. An, J. Zhao, S. Zhao, H. Lv, S. Wang, Drug-Target Interaction Prediction Based on Multisource Information Weighted Fusion, Contrast Media Mol. Imaging 2021 (2021) 6044256. https://doi.org/https://doi.org/10.1155/2021/6044256.
[77] H. Cho, B. Berger, J. Peng, Diffusion Component Analysis: Unraveling Functional Topology in Biological Networks BT - Research in Computational Molecular Biology, in: T.M. Przytycka (Ed.), Springer International Publishing, Cham, 2015: pp. 62–64.
[78] M. Hattori, Y. Okuno, S. Goto, M. Kanehisa, Development of a Chemical Structure Comparison Method for Integrated Analysis of Chemical and Genomic Information in the Metabolic Pathways, J. Am. Chem. Soc.

125 (2003) 11853–11865. https://doi.org/10.1021/ja036030u.
[79] T. van Laarhoven, E. Marchiori, Predicting Drug-Target Interactions for New Drug Compounds Using a Weighted Nearest Neighbor Profile, PLoS One 8 (2013) e66952. https://doi.org/10.1371/journal.pone.0066952.
[80] J.-P. Mei, C.-K. Kwoh, P. Yang, X.-L. Li, J. Zheng, Drug–target interaction prediction by learning from local information and neighbors, Bioinformatics 29 (2013) 238–245. https://doi.org/10.1093/bioinformatics/bts670.
[81] K.-I. Goh, M.E. Cusick, D. Valle, B. Childs, M. Vidal, A.-L. Barabási, The human disease network, Proc. Natl. Acad. Sci. 104 (2007) 8685–8690. https://doi.org/10.1073/pnas.0701361104.
[82] F. Belleau, M.-A. Nolin, N. Tourigny, P. Rigault, J. Morissette, Bio2RDF: Towards a mashup to build bioinformatics knowledge systems, J. Biomed. Inform. 41 (2008) 706–716. https://doi.org/https://doi.org/10.1016/j.jbi.2008.03.004.
[83] T.U. Consortium, The Universal Protein Resource (UniProt), Nucleic Acids Res. 36 (2008) D190–D195. https://doi.org/10.1093/nar/gkm895.
[84] S. Povey, R. Lovering, E. Bruford, M. Wright, M. Lush, H. Wain, The HUGO Gene Nomenclature Committee (HGNC), Hum. Genet. 109 (2001) 678–680. https://doi.org/10.1007/s00439-001-0615-0.
[85] A. Hamosh, A.F. Scott, J.S. Amberger, C.A. Bocchini, V.A. McKusick, Online Mendelian Inheritance in Man (OMIM), a knowledgebase of human genes and genetic disorders, Nucleic Acids Res. 33 (2005) D514–D517. https://doi.org/10.1093/nar/gki033.
[86] M. Kanehisa, S. Goto, KEGG: Kyoto Encyclopedia of Genes and Genomes, Nucleic Acids Res. 28 (2000) 27–30. https://doi.org/10.1093/nar/28.1.27.
[87] E.E. Bolton, Y. Wang, P.A. Thiessen, S.H. Bryant, Chapter 12 - PubChem: Integrated Platform of Small Molecules and Biological Activities, in: R.A. Wheeler, D.C.B.T.-A.R. in C.C. Spellmeyer (Eds.), Elsevier, 2008: pp. 217–241. https://doi.org/https://doi.org/10.1016/S1574-1400(08)00012-1.
[88] O. Bodenreider, The Unified Medical Language System (UMLS): integrating biomedical terminology, Nucleic Acids Res. 32 (2004) D267–D270. https://doi.org/10.1093/nar/gkh061.
[89] C.E. Lipscomb, Medical Subject Headings (MeSH)., Bull. Med. Libr. Assoc. 88 (2000) 265–266.
[90] H. Yang, C. Qin, Y.H. Li, L. Tao, J. Zhou, C.Y. Yu, F. Xu, Z. Chen, F. Zhu, Y.Z. Chen, Therapeutic target database update 2016: enriched resource for bench to clinical drug target and targeted pathway information, Nucleic Acids Res. 44 (2016) D1069–D1074. https://doi.org/10.1093/nar/gkv1230.
[91] A. Gaulton, L.J. Bellis, A.P. Bento, J. Chambers, M. Davies, A. Hersey, Y. Light, S. McGlinchey, D. Michalovich, B. Al-Lazikani, J.P. Overington, ChEMBL: a large-scale bioactivity database for drug discovery, Nucleic Acids Res. 40 (2012) D1100–D1107. https://doi.org/10.1093/nar/gkr777.
[92] T. Liu, Y. Lin, X. Wen, R.N. Jorissen, M.K. Gilson, BindingDB: a web-accessible database of experimentally determined protein–ligand binding affinities, Nucleic Acids Res. 35 (2007) D198–D201. https://doi.org/10.1093/nar/gkl999.
[93] A.J. Pawson, J.L. Sharman, H.E. Benson, E. Faccenda, S.P.H. Alexander, O.P. Buneman, A.P. Davenport, J.C. McGrath, J.A. Peters, C. Southan, M. Spedding, W. Yu, A.J. Harmar, NC-IUPHAR, The IUPHAR/BPS Guide to PHARMACOLOGY: an expert-driven knowledgebase of drug targets and their ligands, Nucleic Acids Res. 42 (2014) D1098–D1106. https://doi.org/10.1093/nar/gkt1143.
[94] H. Wang, F. Huang, Z. Xiong, W. Zhang, A heterogeneous network-based method with attentive meta-path extraction for predicting drug–target interactions, Brief. Bioinform. 23 (2022) bbac184. https://doi.org/10.1093/bib/bbac184.
[95] T. Nguyen, H. Le, T.P. Quinn, T. Nguyen, T.D. Le, S. Venkatesh, GraphDTA: predicting drug–target binding affinity with graph neural networks, Bioinformatics 37 (2021) 1140–1147. https://doi.org/10.1093/bioinformatics/btaa921.
[96] B. Ramsundar, P. Eastman, P. Walters, V. Pande, Deep Learning for the Life Sciences: Applying Deep Learning to Genomics, Microscopy, Drug Discovery, and More, O'Reilly Media, 2019. https://books.google.com/books?id=5uiRDwAAQBAJ.
[97] K. Xu, W. Hu, J. Leskovec, S. Jegelka, How powerful are graph neural networks?, ArXiv Prepr. ArXiv1810.00826 (2018).
[98] P. Bai, F. Miljković, B. John, H. Lu, Interpretable bilinear attention network with domain adaptation improves drug–target prediction, Nat. Mach. Intell. 5 (2023) 126–136. https://doi.org/10.1038/s42256-022-00605-1.
[99] M. Gao, D. Zhang, Y. Chen, Y. Zhang, Z. Wang, X. Wang, S. Li, Y. Guo, G.I. Webb, A.T.N. Nguyen, L. May, J. Song, GraphormerDTI: A graph transformer-based approach for drug-target interaction prediction,

Comput. Biol. Med. 173 (2024) 108339. https://doi.org/https://doi.org/10.1016/j.compbiomed.2024.108339.
[100] M. Jiang, Z. Li, S. Zhang, S. Wang, X. Wang, Q. Yuan, Z. Wei, Drug–target affinity prediction using graph neural network and contact maps, RSC Adv. 10 (2020) 20701–20712. https://doi.org/10.1039/D0RA02297G.
[101] M. Michel, D. Menéndez Hurtado, A. Elofsson, PconsC4: fast, accurate and hassle-free contact predictions, Bioinformatics 35 (2018) 2677–2679. https://doi.org/10.1093/bioinformatics/bty1036.
[102] Y. Liu, L. Xing, L. Zhang, H. Cai, M. Guo, GEFormerDTA: drug target affinity prediction based on transformer graph for early fusion, Sci. Rep. 14 (2024) 7416. https://doi.org/10.1038/s41598-024-57879-1.
[103] T.M. Nguyen, T. Nguyen, T.M. Le, T. Tran, GEFA: Early Fusion Approach in Drug-Target Affinity Prediction, IEEE/ACM Trans. Comput. Biol. Bioinforma. 19 (2022) 718–728. https://doi.org/10.1109/TCBB.2021.3094217.
[104] H. Qi, T. Yu, W. Yu, C. Liu, Drug–target affinity prediction with extended graph learning-convolutional networks, BMC Bioinformatics 25 (2024) 75. https://doi.org/10.1186/s12859-024-05698-6.
[105] Y. Luo, Y. Liu, J. Peng, Calibrated geometric deep learning improves kinase–drug binding predictions, Nat. Mach. Intell. 5 (2023) 1390–1401. https://doi.org/10.1038/s42256-023-00751-0.
[106] Y. Shi, Z. Huang, S. Feng, H. Zhong, W. Wang, Y. Sun, Masked label prediction: Unified message passing model for semi-supervised classification, ArXiv Prepr. ArXiv2009.03509 (2020).
[107] B. Jing, S. Eismann, P. Suriana, R.J.L. Townshend, R. Dror, Learning from protein structure with geometric vector perceptrons, ArXiv Prepr. ArXiv2009.01411 (2020).
[108] H. Wu, J. Liu, T. Jiang, Q. Zou, S. Qi, Z. Cui, P. Tiwari, Y. Ding, AttentionMGT-DTA: A multi-modal drug-target affinity prediction using graph transformer and attention mechanism, Neural Networks 169 (2024) 623–636. https://doi.org/https://doi.org/10.1016/j.neunet.2023.11.018.
[109] J. Jumper, R. Evans, A. Pritzel, T. Green, M. Figurnov, O. Ronneberger, K. Tunyasuvunakool, R. Bates, A. Žídek, A. Potapenko, A. Bridgland, C. Meyer, S.A.A. Kohl, A.J. Ballard, A. Cowie, B. Romera-Paredes, S. Nikolov, R. Jain, J. Adler, T. Back, S. Petersen, D. Reiman, E. Clancy, M. Zielinski, M. Steinegger, M. Pacholska, T. Berghammer, S. Bodenstein, D. Silver, O. Vinyals, A.W. Senior, K. Kavukcuoglu, P. Kohli, D. Hassabis, Highly accurate protein structure prediction with AlphaFold, Nature 596 (2021) 583–589. https://doi.org/10.1038/s41586-021-03819-2.
[110] Z. Lin, H. Akin, R. Rao, B. Hie, Z. Zhu, W. Lu, N. Smetanin, R. Verkuil, O. Kabeli, Y. Shmueli, A. dos Santos Costa, M. Fazel-Zarandi, T. Sercu, S. Candido, A. Rives, Evolutionary-scale prediction of atomic-level protein structure with a language model, Science (80-. ). 379 (2023) 1123–1130. https://doi.org/10.1126/science.ade2574.
[111] S.M. Marinka Zitnik Rok Sosič, J. Leskovec, {BioSNAP Datasets}: {Stanford} Biomedical Network Dataset Collection, (2018).
[112] M. Tsubaki, K. Tomii, J. Sese, Compound–protein interaction prediction with end-to-end learning of neural networks for graphs and sequences, Bioinformatics 35 (2019) 309–318. https://doi.org/10.1093/bioinformatics/bty535.
[113] J.T. Metz, E.F. Johnson, N.B. Soni, P.J. Merta, L. Kifle, P.J. Hajduk, Navigating the kinome, Nat. Chem. Biol. 7 (2011) 200–202. https://doi.org/10.1038/nchembio.530.
[114] Q. Feng, E. Dueva, A. Cherkasov, M. Ester, Padme: A deep learning-based framework for drug-target interaction prediction, ArXiv Prepr. ArXiv1807.09741 (2018).
[115] B.-W. Zhao, X.-R. Su, P.-W. Hu, Y.-A. Huang, Z.-H. You, L. Hu, iGRLDTI: an improved graph representation learning method for predicting drug–target interactions over heterogeneous biological information network, Bioinformatics 39 (2023) btad451. https://doi.org/10.1093/bioinformatics/btad451.
[116] W. Zhang, M. Yang, Z. Sheng, Y. Li, W. Ouyang, Y. Tao, Z. Yang, B. CUI, Node Dependent Local Smoothing for Scalable Graph Learning, in: M. Ranzato, A. Beygelzimer, Y. Dauphin, P.S. Liang, J.W. Vaughan (Eds.), Adv. Neural Inf. Process. Syst., Curran Associates, Inc., 2021: pp. 20321–20332. https://proceedings.neurips.cc/paper_files/paper/2021/file/a9eb812238f753132652ae09963a05e9-Paper.pdf.
[117] Z. Cheng, D. Xu, D. Ding, Y. Ding, Prediction of Drug-Target Interactions With High- Quality Negative Samples and a Network-Based Deep Learning Framework, IEEE J. Biomed. Heal. Informatics 29 (2025) 1567–1578. https://doi.org/10.1109/JBHI.2024.3354953.
[118] J. Shen, J. Zhang, X. Luo, W. Zhu, K. Yu, K. Chen, Y. Li, H. Jiang, Predicting protein–protein interactions based only on sequences information, Proc. Natl. Acad. Sci. 104 (2007) 4337–4341. https://doi.org/10.1073/pnas.0607879104.
[119] B. Yang, Y. Liu, J. Wu, F. Bai, M. Zheng, J. Zheng, GENNDTI: Drug-Target Interaction Prediction Using Graph Neural Network Enhanced by Router Nodes, IEEE J. Biomed. Heal. Informatics 28 (2024) 7588–

7598. https://doi.org/10.1109/JBHI.2024.3402529.
[120] T. Zhao, Y. Hu, L.R. Valsdottir, T. Zang, J. Peng, Identifying drug–target interactions based on graph convolutional network and deep neural network, Brief. Bioinform. 22 (2021) 2141–2150. https://doi.org/10.1093/bib/bbaa044.
[121] U. Veleiro, J. de la Fuente, G. Serrano, M. Pizurica, M. Casals, A. Pineda-Lucena, S. Vicent, I. Ochoa, O. Gevaert, M. Hernaez, GeNNius: an ultrafast drug–target interaction inference method based on graph neural networks, Bioinformatics 40 (2024) btad774. https://doi.org/10.1093/bioinformatics/btad774.
[122] J. Gasteiger, A. Bojchevski, S. Günnemann, Predict then propagate: Graph neural networks meet personalized pagerank, ArXiv Prepr. ArXiv1810.05997 (2018).
[123] W. Wang, S. Yang, J. Li, Drug target predictions based on heterogeneous graph inference., in: Pacific Symp. Biocomput., World Scientific, 2013: pp. 53–64.
[124] W. Ba-alawi, O. Soufan, M. Essack, P. Kalnis, V.B. Bajic, DASPfind: new efficient method to predict drug–target interactions, J. Cheminform. 8 (2016) 15. https://doi.org/10.1186/s13321-016-0128-4.
[125] D. Binns, E. Dimmer, R. Huntley, D. Barrell, C. O'Donovan, R. Apweiler, QuickGO: a web-based tool for Gene Ontology searching, Bioinformatics 25 (2009) 3045–3046. https://doi.org/10.1093/bioinformatics/btp536.
[126] M. Gillespie, B. Jassal, R. Stephan, M. Milacic, K. Rothfels, A. Senff-Ribeiro, J. Griss, C. Sevilla, L. Matthews, C. Gong, C. Deng, T. Varusai, E. Ragueneau, Y. Haider, B. May, V. Shamovsky, J. Weiser, T. Brunson, N. Sanati, L. Beckman, X. Shao, A. Fabregat, K. Sidiropoulos, J. Murillo, G. Viteri, J. Cook, S. Shorser, G. Bader, E. Demir, C. Sander, R. Haw, G. Wu, L. Stein, H. Hermjakob, P. D'Eustachio, The reactome pathway knowledgebase 2022, Nucleic Acids Res. 50 (2022) D687–D692. https://doi.org/10.1093/nar/gkab1028.
[127] M. Martens, A. Ammar, A. Riutta, A. Waagmeester, D.N. Slenter, K. Hanspers, R. A. Miller, D. Digles, E.N. Lopes, F. Ehrhart, L.J. Dupuis, L.A. Winckers, S.L. Coort, E.L. Willighagen, C.T. Evelo, A.R. Pico, M. Kutmon, WikiPathways: connecting communities, Nucleic Acids Res. 49 (2021) D613–D621. https://doi.org/10.1093/nar/gkaa1024.
[128] D. Szklarczyk, A.L. Gable, D. Lyon, A. Junge, S. Wyder, J. Huerta-Cepas, M. Simonovic, N.T. Doncheva, J.H. Morris, P. Bork, L.J. Jensen, C. von Mering, STRING v11: protein–protein association networks with increased coverage, supporting functional discovery in genome-wide experimental datasets, Nucleic Acids Res. 47 (2019) D607–D613. https://doi.org/10.1093/nar/gky1131.
[129] S. Yun, M. Jeong, R. Kim, J. Kang, H.J. Kim, Graph Transformer Networks, in: H. Wallach, H. Larochelle, A. Beygelzimer, F. d\textquotesingle Alché-Buc, E. Fox, R. Garnett (Eds.), Adv. Neural Inf. Process. Syst., Curran Associates, Inc., 2019. https://proceedings.neurips.cc/paper_files/paper/2019/file/9d63484abb477c97640154d40595a3bb-Paper.pdf.
[130] C. Shi, Y. Li, J. Zhang, Y. Sun, P.S. Yu, A survey of heterogeneous information network analysis, IEEE Trans. Knowl. Data Eng. 29 (2017) 17–37. https://doi.org/10.1109/TKDE.2016.2598561.
[131] M. Li, X. Cai, S. Xu, H. Ji, Metapath-aggregated heterogeneous graph neural network for drug–target interaction prediction, Brief. Bioinform. 24 (2023) bbac578. https://doi.org/10.1093/bib/bbac578.
[132] G. Fu, Y. Ding, A. Seal, B. Chen, Y. Sun, E. Bolton, Predicting drug target interactions using meta-path-based semantic network analysis, BMC Bioinformatics 17 (2016) 160. https://doi.org/10.1186/s12859-016-1005-x.
[133] H. Shi, S. Liu, J. Chen, X. Li, Q. Ma, B. Yu, Predicting drug-target interactions using Lasso with random forest based on evolutionary information and chemical structure, Genomics 111 (2019) 1839–1852. https://doi.org/https://doi.org/10.1016/j.ygeno.2018.12.007.
[134] K. Huang, C. Xiao, L.M. Glass, M. Zitnik, J. Sun, SkipGNN: predicting molecular interactions with skip-graph networks, Sci. Rep. 10 (2020) 21092. https://doi.org/10.1038/s41598-020-77766-9.
[135] Y. Li, G. Qiao, X. Gao, G. Wang, Supervised graph co-contrastive learning for drug–target interaction prediction, Bioinformatics 38 (2022) 2847–2854. https://doi.org/10.1093/bioinformatics/btac164.
[136] Q. Ning, Y. Wang, Y. Zhao, J. Sun, L. Jiang, K. Wang, M. Yin, DMHGNN: Double multi-view heterogeneous graph neural network framework for drug-target interaction prediction, Artif. Intell. Med. 159 (2025) 103023. https://doi.org/https://doi.org/10.1016/j.artmed.2024.103023.
[137] J. Li, J. Wang, H. Lv, Z. Zhang, Z. Wang, IMCHGAN: Inductive Matrix Completion With Heterogeneous Graph Attention Networks for Drug-Target Interactions Prediction, IEEE/ACM Trans. Comput. Biol. Bioinforma. 19 (2022) 655–665. https://doi.org/10.1109/TCBB.2021.3088614.
[138] Y. Su, Z. Hu, F. Wang, Y. Bin, C. Zheng, H. Li, H. Chen, X. Zeng, AMGDTI: drug–target interaction

prediction based on adaptive meta-graph learning in heterogeneous network, Brief. Bioinform. 25 (2024) bbad474. https://doi.org/10.1093/bib/bbad474.
[139] F. Wu, A. Souza, T. Zhang, C. Fifty, T. Yu, K. Weinberger, Simplifying Graph Convolutional Networks, in: K. Chaudhuri, R. Salakhutdinov (Eds.), Proc. 36th Int. Conf. Mach. Learn., PMLR, 2019: pp. 6861–6871. https://proceedings.mlr.press/v97/wu19e.html.
[140] Z. Liu, Q. Chen, W. Lan, H. Pan, X. Hao, S. Pan, GADTI: Graph Autoencoder Approach for DTI Prediction From Heterogeneous Network, Front. Genet. 12 (2021). https://www.frontiersin.org/journals/genetics/articles/10.3389/fgene.2021.650821.
[141] B. Yang, W. Yih, X. He, J. Gao, L. Deng, Embedding entities and relations for learning and inference in knowledge bases, ArXiv Prepr. ArXiv1412.6575 (2014).
[142] L. Jiang, J. Sun, Y. Wang, Q. Ning, N. Luo, M. Yin, Identifying drug–target interactions via heterogeneous graph attention networks combined with cross-modal similarities, Brief. Bioinform. 23 (2022) bbac016. https://doi.org/10.1093/bib/bbac016.
[143] Y. Zheng, H. Peng, X. Zhang, X. Gao, J. Li, Predicting Drug Targets from Heterogeneous Spaces using Anchor Graph Hashing and Ensemble Learning, in: 2018 Int. Jt. Conf. Neural Networks, 2018: pp. 1–7. https://doi.org/10.1109/IJCNN.2018.8489028.
[144] N. Zong, N. Li, A. Wen, V. Ngo, Y. Yu, M. Huang, S. Chowdhury, C. Jiang, S. Fu, R. Weinshilboum, G. Jiang, L. Hunter, H. Liu, BETA: a comprehensive benchmark for computational drug–target prediction, Brief. Bioinform. 23 (2022) bbac199. https://doi.org/10.1093/bib/bbac199.
[145] J. de la Fuente, G. Serrano, U. Veleiro, M. Casals, L. Vera, M. Pizurica, N. Gómez-Cebrián, L. Puchades-Carrasco, A. Pineda-Lucena, I. Ochoa, S. Vicent, O. Gevaert, M. Hernaez, Towards a more inductive world for drug repurposing approaches, Nat. Mach. Intell. 7 (2025) 495–508. https://doi.org/10.1038/s42256-025-00987-y.
[146] M. Bechler-Speicher, A. Globerson, R. Gilad-Bachrach, The Intelligible and Effective Graph Neural Additive Network, in: A. Globerson, L. Mackey, D. Belgrave, A. Fan, U. Paquet, J. Tomczak, C. Zhang (Eds.), Adv. Neural Inf. Process. Syst., Curran Associates, Inc., 2024: pp. 90552–90578. https://proceedings.neurips.cc/paper_files/paper/2024/file/a4c3a66ed818455b8bbe591b6a5d0f56-Paper-Conference.pdf.
[147] Z. Tanoli, S. Aron, T. and Aittokallio, Validation guidelines for drug-target prediction methods, Expert Opin. Drug Discov. 20 (2025) 31–45. https://doi.org/10.1080/17460441.2024.2430955.
[148] A.N. Angelopoulos, S. Bates, Conformal Prediction: A Gentle Introduction, Found. Trends® Mach. Learn. 16 (2023) 494–591. https://doi.org/10.1561/2200000101.
[149] J. Abramson, J. Adler, J. Dunger, R. Evans, T. Green, A. Pritzel, O. Ronneberger, L. Willmore, A.J. Ballard, J. Bambrick, S.W. Bodenstein, D.A. Evans, C.-C. Hung, M. O'Neill, D. Reiman, K. Tunyasuvunakool, Z. Wu, A. Žemgulytė, E. Arvaniti, C. Beattie, O. Bertolli, A. Bridgland, A. Cherepanov, M. Congreve, A.I. Cowen-Rivers, A. Cowie, M. Figurnov, F.B. Fuchs, H. Gladman, R. Jain, Y.A. Khan, C.M.R. Low, K. Perlin, A. Potapenko, P. Savy, S. Singh, A. Stecula, A. Thillaisundaram, C. Tong, S. Yakneen, E.D. Zhong, M. Zielinski, A. Žídek, V. Bapst, P. Kohli, M. Jaderberg, D. Hassabis, J.M. Jumper, Accurate structure prediction of biomolecular interactions with AlphaFold 3, Nature 630 (2024) 493–500. https://doi.org/10.1038/s41586-024-07487-w.